\documentclass[aps,prl,twocolumn,superscriptaddress]{revtex4-1}
\usepackage[utf8]{inputenc}
\usepackage{amsthm,amssymb,amsmath}
\usepackage{color}
\usepackage{graphicx,xcolor}
\usepackage{bbold}
\usepackage[colorlinks=true,allcolors=blue]{hyperref}

\usepackage{extarrows}
\usepackage{microtype}
\usepackage{siunitx}
\usepackage{float}
\usepackage{booktabs}
\usepackage{xcolor}
\usepackage{mathtools}
\usepackage[normalem]{ulem}

\begin{document}
\title{A purified input-output pseudomode model for structured open quantum systems}

\author{Pengfei Liang}
\email{pfliang@imu.edu.cn}
\affiliation{Center for Quantum Physics and Technologies, School of Physical Science and Technology, Inner Mongolia University, Hohhot 010021, China}
\affiliation{Graduate School of China Academy of Engineering Physics, Haidian District, Beijing, 100193, China}

\author{Neill Lambert}
\email{nwlambert@gmail.com}
\affiliation{Theoretical Physics Laboratory, Cluster for Pioneering Research, RIKEN, Wakoshi, Saitama 351-0198, Japan}
\affiliation{Quantum Computing Center, RIKEN, Wakoshi, Saitama, 351-0198, Japan}

\author{Si Luo}
\affiliation{Department of Chemistry, Fudan University, Shanghai 200433, P. R. China}

\author{Lingzhen Guo}
\affiliation{Center for Joint Quantum Studies and Department of Physics, School of Science, Tianjin University, Tianjin 300072, China}

\author{Mauro Cirio}
\email{cirio.mauro@gmail.com}
\affiliation{Graduate School of China Academy of Engineering Physics, Haidian District, Beijing, 100193, China}

\date{\today}
\begin{abstract}
A full understanding of open quantum systems requires the  characterization of both system and environmental properties. However, the complexity of the environmental statistics in the presence of strong system-bath hybridization and long memory effects usually prevents effective non-perturbative methods from going beyond the analysis of the reduced system dynamics. Here we present a model consisting of purified auxiliary bosonic modes to describe, alongside properties of the system, the dynamics of  environmental observables for bosonic baths prepared in non-Gaussian initial states. We numerically exemplify this method by simulating non-Markovian multi-photon transfer processes on a coupled cavity waveguide system in the large time delay regime.  
\end{abstract}

\maketitle

\emph{Introduction.}
Quantum environments are not simply a
source of dissipation and decoherence~\cite{Petruccione,Gardiner}, but they can also play a constructive role in mediating interactions between distant nodes of quantum networks~\cite{https://doi.org/10.1002/lpor.202100219}, or in achieving selective directionality in quantum information transfer~\cite{RevModPhys.95.015002}. Alongside 
the simulation of reduced system dynamics, which has been the main focus of the theory of open quantum systems~\cite{Petruccione,Gardiner}, the analysis 
of environmental information becomes increasingly important due to advances in quantum technologies~\cite{Acin_2018}. While in the Markovian regime, the well-established input-output relations allow to fully track environmental properties, it is not clear whether similar results can be achieved in the presence of strong system-bath hybridization~\cite{FriskKockum2019,RevModPhys.91.025005} or large time delays~\cite{PhysRevA.90.013837,PhysRevLett.120.140404,Andersson2019,GuoOBS,PhysRevA.104.053701}. 

To approach this challenge, it is possible to directly evolve the many-body wavefunction of the original continuum model 
by discretizing environmental degrees of freedom~\cite{PhysRevB.92.155126,10.1063/1.5135363,Shuai_review}, or to solve unitarily equivalent models obtained from the
reaction coordinate mapping~\cite{Garg,Martinazzo,iles2014environmental,Woods,PhysRevB.97.205405,Melina}, the chain mapping~\cite{Chin_2010,PhysRevLett.105.050404,PhysRevLett.123.090402,PhysRevB.101.155134}, or the polaron transformation~\cite{Holstein1,Holstein2,Jackson,Silbey,Silbey2,PhysRevB.57.347,PhysRevB.65.235311,Jang,Jang2,PhysRevLett.103.146404}. On the other hand, effective methods such as the hierarchical equations of motion (HEOM)~\cite{Tanimura_3,Tanimura_1,Ishizaki_1,Tanimura_2,Ishizaki_2,PhysRevLett.104.250401,PhysRevLett.109.266403,PhysRevA.85.062323,Tanimura_2014,Song_Shi,FreePoles,Dan_Shi,xu2023universalframeworkquantumdissipationminimally,Tanimura_2020,Tanimura_2021}, dissipatons~\cite{Yan_1,10.1063/1.4905494,Yan_2,Yan_3,Yan_4,10.1063/5.0123999,10.1063/5.0155585,10.1063/5.0151239,Chen_Yan,li2024quantumsimulationnonmarkovianopen}, and pseudomodes~\cite{garraway1997,PhysRevLett.110.086403,PhysRevB.89.165105,PhysRevB.92.245125,Schwarz,Dorda,Mascherpa,Tamascelli,Lambert2019,PhysRevResearch.2.043058,PhysRevA.101.052108,LuoSiPRXQ,albarelli2024,menczel2024nonhermitian,PhysRevA.110.022221,PhysRevLett.126.093601,PhysRevLett.132.106902,gunhee2024prb}, 
are developed to model the effects of the external continuum using auxiliary degrees of freedom for efficient simulation of the reduced system dynamics. However, these effective methods only allow to access bath properties related to system correlations~\cite{PhysRevLett.126.093601} or bath coupling operators~\cite{PhysRevResearch.5.013181,Song_Shi}. The same limitation exists for the recently proposed tensor-network algorithms for the calculation of the Feymann-Vernon influence functional~\cite{Strathearn_2017,Strathearn2018,PhysRevLett.123.240602,PhysRevLett.129.173001,PhysRevLett.126.200401,PhysRevLett.128.167403,PRXQuantum.3.010321,PhysRevLett.132.200403,PhysRevX.14.011010}. 

\begin{figure}
\includegraphics[clip,width=8.5cm]{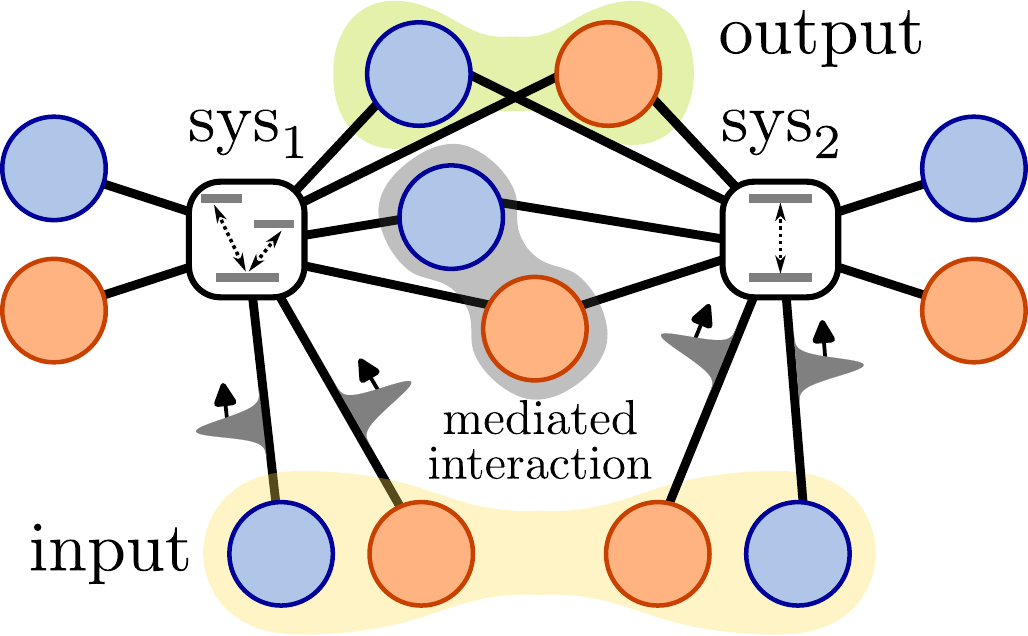}%
\caption{Sketch of a purified pseudomode model to simulate the effects of an environment mediating the interaction between two quantum systems $\text{sys}_{1}$ and $\text{sys}_2$. Modes in the green region allow for the read-out of bath observables while those in the yellow region encode the preparation of the bath in a non-Gaussian state. The remaining modes reproduce the reduced system dynamics, including the bath-mediated interaction modeled by modes in the gray region. Purified pseudomodes in blue (orange) are defined to model bath correlations forward (backward) in time, see Fig.~\ref{fig:continuation}. 
}\label{fig:schematic}
\end{figure}

\begin{figure}
\includegraphics[clip,width=8.5cm]{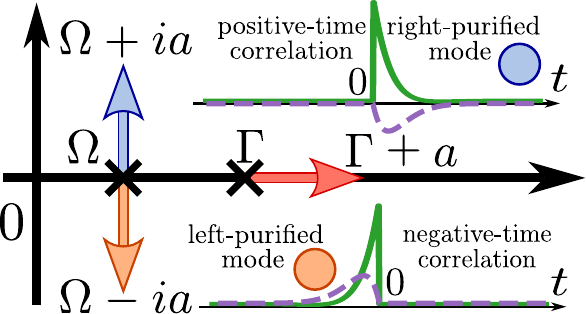}
\caption{Analytical continuation of the frequency $\Omega$ and decay rate $\Gamma$ of a pseudomode along the paths $\Omega\to\Omega\pm ia$ for $\Gamma\to\Gamma+a$ in the complex plane. $\mathcal{C}_\text{PPM}^\pm(t)$ in Eq.~(\ref{eq:Cppm}) are illustrated in the insets by their real (solid) and imaginary (dashed) parts.
}\label{fig:continuation}
\end{figure}

In this work, we introduce a method based on the pseudomode theory~\cite{garraway1997,Tamascelli,Lambert2019} for numerically exact description of the reduced system dynamics and generic bath observables, which is valid even in the presence of non-Gaussian environmental initial states. We achieve this by constructing effective models (see Fig.~\ref{fig:schematic}) made of a new class of auxiliary bosonic modes, which are defined by analytically continuing some of the pseudomode parameters (see Fig.~\ref{fig:continuation}) to allow their state to remain pure during the whole dynamics. For this reason we call them {\it purified pseudomodes}. In addition, our method also reveals the explicit connection between the pseudomode theory and the HEOM~\cite{FreePoles,Yan_1,10.1063/1.4905494,Yan_2,Yan_3,Yan_4,10.1063/5.0123999,10.1063/5.0155585,10.1063/5.0151239,Chen_Yan,li2024quantumsimulationnonmarkovianopen}, thereby integrating the physically intuitive characteristics of the former with the efficiency of the latter. As an illustration, we show multi-photon transfer processes in a coupled cavity waveguide model~\cite{PhysRevLett.98.083603,Yao:09,Yu:21} in the deep non-Markovian regime with large time delays. 

\emph{System dynamics and environmental correlations.}
We start by deﬁning our open setting~\cite{Petruccione,Gardiner}, which includes a system S with Hamiltonian $H_\text{S}$ and a bath B with Hamiltonian $H_\text{B}=\sum_k\omega_k b_k^\dagger b_k$ describing a collection of independent harmonic modes $b_k$ having frequency $\omega_k$. The system and bath is coupled via the system-bath interaction $H_\text{SB} = \sum_{n=1}^N S_n X_n$, where $S_n$ and $X_n$ represent the system and bath coupling operators, respectively. Here, we assume $X_n$ to be linear combinations of the ladder operators $b_k$ and $b_k^\dagger$. We introduce the reduced density matrix $\rho_\text{S}(t)=\text{Tr}_\text{B}\rho(t)$ by tracing out the environmental degrees of freedom in the full density matrix $\rho(t)$.
We further assume that the system and the bath are initially independent, i.e, 
$\rho(0)=\rho_\text{S}\otimes\rho_\text{B}$ for a generic system state $\rho_\text{S}$ and a Gaussian bath state $\rho_\text{B}$. 

The Gaussian statistics encoded in $\rho_\text{B}$ and the linearity of $X_n$ allow to write an explicit expression for $\rho_\text{S}(t)$ in the interaction picture as  
$\rho_\text{S}(t) =\mathcal{T}_\text{S}\;\exp{[-\mathcal{F}_t]}\rho_\text{S}$,  
where $\mathcal{F}_t$ is an influence superoperator written as~\cite{Petruccione,Aurell,PhysRevB.105.035121}
\begin{equation}\label{eq:Ft}
\mathcal{F}_t = \sum_{\substack{n,m \\ \alpha,\beta\in\{l,r\}}}\int_0^{t}d\tau_2\int_0^{\tau_2}d\tau_1  \mathcal{C}^{\alpha,\beta}_{n,m}(\tau_2,\tau_1)\mathcal{S}_n^\alpha(\tau_2)\mathcal{S}_m^\beta(\tau_1), 
\end{equation}
i.e., as a function of the correlation matrix $\mathcal{C}^{\alpha,\beta}_{n,m}(\tau_2,\tau_1)=\langle \mathcal{T}_\text{B}\mathcal{X}_n^\alpha(\tau_2)\mathcal{X}_m^\beta(\tau_1)\rangle_0$ with $\langle\boldsymbol{\cdot}\rangle_0\equiv \text{Tr}[\boldsymbol{\cdot}\rho_\text{B}]$ and $\alpha,\beta\in\{l,r\}$. The compactness of this expression originates from defining $\mathcal{S}_n^\alpha$ and $\mathcal{X}_n^\alpha$ as the superoperators required to reproduce the commutators involving $H_\text{SB}$ in the Dyson series, which is achieved by defining the left and right components as
$\mathcal{S}_n^l[\cdot] = S_n[\cdot]$, $\mathcal{S}_n^r[\cdot] = -[\cdot]S_n$, $\mathcal{X}_n^l[\cdot]=X_n[\cdot]$, and $\mathcal{X}_n^r=[\cdot]X_n$.  The operators $\mathcal{T}_\text{B}$ and $\mathcal{T}_\text{S}$ ensure the time-ordering of the superoperators $\mathcal{X}_n^\alpha$ and $\mathcal{S}_n^\alpha$, respectively.

In order to analyze environmental properties other than the reduced system dynamics, we consider general correlations of the form \cite{cirio2024inputoutputhierarchicalequationsmotion}
\begin{equation}\label{eq:Ecorr}
\rho_{\text{S};\{t_j\}}^{\{\alpha_j\}}(t) \equiv\text{Tr}_\text{B}\left[\mathcal{T}_\text{S}\mathcal{T}_\text{B}\Phi_M^{\alpha_M}(t_M)\cdots\Phi_1^{\alpha_1}(t_1)\rho(t)\right], 
\end{equation}
involving $M$ field superoperators $\Phi_j^{\alpha_j}(t_j)$ evaluated at the time $t_j$ in the interaction picture, with $j\in\{1,\cdots,M\}$. Here, each $\Phi_j^{\alpha_j}$ corresponds to the left ($\alpha_j=l$) or right ($\alpha_j=r$) action of a field operator $\phi_j$ linear in $b_k$ and $b_k^\dagger$, i.e.,  $\Phi_j^l(t_j)[\cdot] = \phi_j(t_j)[\cdot]$ and $\Phi_j^r(t_j)[\cdot] = [\cdot]\phi_j(t_j)$.  

As in the analysis of $\rho_\text{S}(t)$, these assumptions allow to use Wick's theorem to write Eq.~(\ref{eq:Ecorr}) as a weighted sum of the matrices~\cite{cirio2024inputoutputhierarchicalequationsmotion} 
\begin{equation}\label{eq:Sat}
 \rho_\text{S}^{\mathfrak{a}}(t) =\mathcal{T}_\text{S}\prod_{j\in \mathfrak{a}} \sum_{n,\alpha}\int_0^t d\tau \,\tilde{\mathcal{C}}_{j,n}^{\alpha_j,\alpha}(t_j,\tau)\mathcal{S}_n^{\alpha}(\tau)\rho_\text{S}(t), 
\end{equation}
in terms of the cross-correlation matrix $\tilde{\mathcal{C}}_{j,n}^{\alpha_j,\alpha}(t_j,\tau)=\langle\mathcal{T}_\text{B}\Phi_{j}^{\alpha_j}(t_j)\mathcal{X}_n^\alpha(\tau)\rangle_0$ and the index $\mathfrak{a}$ running over all possible subsets of the set 
$\{1,\cdots,M\}$. In particular, when $\mathfrak{a}=\emptyset$, we assume that $\rho_\text{S}^\emptyset(t) = \rho_\text{S}(t)$. In other words, the knowledge of $\rho_\text{S}^\mathfrak{a}(t)$ for all $\mathfrak{a}$ allows a direct computation of $\rho_{\text{S};\{t_j\}}^{\{\alpha_j\}}(t)$ which, in turn, can be used to reproduce bath properties beyond, but also including, the reduced system dynamics. For example, they can be used to model non-Gaussian input states and to compute output bath observables. 

The matrices $\rho_\text{S}^{\mathfrak{a}}(t)$ can be computed by solving an extended version of the HEOM~\cite{cirio2024inputoutputhierarchicalequationsmotion}. Here, we present an alternative method to compute them, based on the pseudomode theory. In particular, we show that, whenever an exponential decomposition of each positive- ($t_r>t_s$) or negative-time ($t_r<t_s$) correlation 
\begin{equation}\label{eq:corrAB}
\langle A(t_r)B(t_s)\rangle_0, 
\end{equation}
with $A,B\in\{X_1,\cdots,X_N,\phi_1,\cdots,\phi_M\}$ and $t_r,t_s\in\{\tau_1,\tau_2,\tau,t_1,\cdots,t_M\}$, is available, $\rho_\text{S}^{\mathfrak{a}}(t)$ can also be obtained by solving a simple Lindblad-like master equation in a Hilbert space with optimized local Hilbert dimension. To achieve this, we first present a purification procedure to optimize the pseudomode Hilbert space, thereby defining purified pseudomodes. Then, we use this result
to explicitly construct an effective purified pseudomode model, as illustrated in Fig.~\ref{fig:schematic}, to compute $\rho_\text{S}^{\mathfrak{a}}(t)$.

\emph{Purification of zero-temperature pseudomodes.}
The conventional pseudomode theory~\cite{garraway1997,PhysRevLett.110.086403,PhysRevB.89.165105,PhysRevB.92.245125,Schwarz,Dorda,Mascherpa,Tamascelli,Lambert2019,PhysRevResearch.2.043058,PhysRevA.101.052108,LuoSiPRXQ,albarelli2024,menczel2024nonhermitian,PhysRevA.110.022221,PhysRevLett.126.093601,PhysRevLett.132.106902} aims to reproduce $\rho_\text{S}(t)$ by establishing an effective dissipative environment made of ancillary bosonic modes, each subject to an additional Markovian channel. In fact, the original environment can be modelled using lossy ancillary modes as long as the correlations of the two environments match. As an illustration, we consider the system-bath interaction $H_\text{SB} = SX$ ($N=1$) such that the dynamics only depends on the single function $\mathcal{C}(t) = \langle X(t) X(0)\rangle_0$. For example, the correlation of a single 
zero-temperature pseudomode $d$ with frequency $\Omega$ and decay rate $\Gamma$ is $\mathcal{C}(t)=\lambda^2 \exp{[-i\Omega t - \Gamma \lvert t\rvert]}$ with $\lambda,~\Omega\in\mathbb{R},~\Gamma\in\mathbb{R}_+$. 
The density operator $\rho_\text{PM}$ in the enlarged Hilbert space, consisting of both the system and the pseudomode $d$, satisfies the following Lindblad equation 
\begin{equation}\label{eq:mePM}
i\frac{d\rho_\text{PM}}{dt} = [H_\text{S},\rho_\text{PM}] + L_\text{PM}~\rho_\text{PM},
\end{equation}
with the dynamical semigroup generator $L_\text{PM}[\boldsymbol{\cdot}] = \left[\Omega d^\dagger d + \lambda S(d+d^\dagger),\boldsymbol{\cdot}\right] + i\Gamma\left(2d[\boldsymbol{\cdot}] d^\dagger - d^\dagger d[\boldsymbol{\cdot}] - [\boldsymbol{\cdot}] d^\dagger d\right)$ 
and the initial state $\rho_\text{PM}(0)=\rho_\text{S}\otimes|0\rangle\langle0|$. 
For more complex correlations, additional ancillas (possibly interacting~\cite{PhysRevA.101.052108,PhysRevLett.126.093601,PhysRevLett.132.106902}) should be considered in the pseudomode model. By solving the corresponding Lindblad dynamics and tracing over the pseudomodes, the reduced system dynamics can be ultimately evaluated, i.e., $\rho_\text{S}(t)=\text{Tr}_\text{PM}[\rho_\text{PM}(t)]$.

The effective models are powerful as they are not bound by physical assumptions. In fact, their solutions can be analytically continued in the model parameters to enlarge predictivity. For example, this  can be used to improve detailed balance~\cite{Lambert2019}, to mitigate non-Markovian noise~\cite{PhysRevResearch.6.033083} and for dissipative state engineering~\cite{lambert2023fixingdetailedbalanceancillabased}.  Here we focus on the frequency $\Omega$ and decay rate $\Gamma$ of the single mode example considered above, and analyze the following analytical continuation paths in the complex plane
\begin{equation}\label{eq:acpaths}
\Omega \to \Omega \pm ia,~\Gamma \to \Gamma+a, 
\end{equation}
in terms of a free energy-parameter 
$a\in[0,+\infty)$. In particular, we are interested in the limit $a\rightarrow+\infty$, as shown in Fig.~\ref{fig:continuation}, which corresponds to a decomposition of $\mathcal{C}(t)$ in 
its positive- and negative-time contributions
\begin{equation}\label{eq:Cppm}
\mathcal{C}^{\pm}_\text{PPM}(t) = \lim_{a\to+\infty} \lambda^2e^{-i(\Omega\pm ia) t-(\Gamma+a)\lvert t\rvert} = \mathcal{C}(t)\Theta(\pm t),   
\end{equation}
where $\Theta(t)$ is the Heaviside step function. 
The reason for considering this limit lies in the properties of the modes corresponding to 
these correlations. 

In the Supplemental Material~\cite{supp_mat}, we show that in the limit $a\to+\infty$, the environmental effects on $\rho_\text{S}(t)$ encoded in $\mathcal{C}(t)$ can be identically reproduced by a Lindblad-like master equation
\begin{equation}\label{eq:onemode_effme}
i\frac{d\rho_\text{eff}}{dt} = [H_\text{S},\rho_\text{eff}] + L_+\rho_\text{eff}+L_-\rho_\text{eff},
\end{equation}
with semigroup generators defined as $L_+[\boldsymbol{\cdot}] = (\Omega-i\Gamma)d_+^\dagger d_+[\boldsymbol{\cdot}]+ \lambda d_+[S,\boldsymbol{\cdot}] + \lambda d_+^\dagger S[\boldsymbol{\cdot}]$ and $L_-[\boldsymbol{\cdot}] = -(\Omega+i\Gamma)[\boldsymbol{\cdot}]d_-^\dagger d_- + \lambda [S,\boldsymbol{\cdot}]d_-^\dagger - \lambda[\boldsymbol{\cdot}]S d_-$  involving the bosonic modes $d_\pm$ having complex frequencies $\pm\Omega-i\Gamma$. The equivalence of Eq.~(\ref{eq:onemode_effme}) and Eq.~(\ref{eq:mePM}) is manifested in the equality $\rho_\text{S}(t) = \text{Tr}_\pm[\rho_\text{eff}(t)] = \text{Tr}_\text{PM}[\rho_\text{PM}(t)]$, with $\text{Tr}_\pm[\cdot]$ denoting the trace over both $d_+$ and $d_-$. Importantly, since $d_\pm$ and $d_\pm^\dagger$ always act on one side in $L_\pm$, the pseudomode state becomes pure, i.e., it can be represented as a state vector instead of a density matrix. In this way, we have thus defined the right/left-purified pseudomodes $d_{+/-}$ characterized by the correlations $\mathcal{C}_\text{PPM}^{+/-}(t)$, see insets of Fig.~\ref{fig:continuation}. 
This purification procedure allows a reduction of the  local Hilbert dimension in Eq.~(\ref{eq:onemode_effme}) which can be further utilized in tensor-network algorithms, see the Supplemental Material~\cite{supp_mat} for a more in-depth analysis of computational complexity.

\emph{Purified pseudomode models for system dynamics and bath input-output.} 
We now use the purified pseudomodes to explicitly define an effective model to compute $\rho_\text{S}^{\mathfrak{a}}(t)$. This is achieved by simply introducing independent 
purified modes to generalize Eq.~(\ref{eq:onemode_effme}) as
\begin{equation}\label{eq:PPMme}
i\frac{d\rho_\text{eff}}{dt} = [H_\text{S},\rho_\text{eff}] 
+ \sum_\alpha L_{\alpha+}\rho_\text{eff} + \sum_\beta L_{\beta-}\rho_\text{eff},   
\end{equation}
and by finding a pseudomode version of the original environmental fields, i.e.,
\begin{equation}\label{eq:Phimapping}
\Phi_j^{\alpha_j} \to \Phi_{\text{PPM},j}^{\alpha_j} = \sum_\alpha \Phi_{\text{PPM},j\alpha+}^{\alpha_j} + \sum_\beta \Phi_{\text{PPM},j\beta-}^{\alpha_j}.
\end{equation}
In other words, the generators $L_{\alpha+}$, $L_{\beta-}$ and the pseudomode field superoperators $\Phi_{\text{PPM},j\alpha+}^{\alpha_j}$ and $\Phi_{\text{PPM},j\beta-}^{\alpha_j}$ are to be determined given the bath correlations in Eq.~(\ref{eq:corrAB}). Then, input-output properties of the original model can be obtained by solving Eq.~(\ref{eq:PPMme}) and taking expectations involving the quantities in Eq.~(\ref{eq:Phimapping}).

While a rigorous analysis of the construction can be found in the Supplemental Material~\cite{supp_mat}, here we present an intuitive and practical solution. In fact, since the purified pseudomodes are initially in their vacuum, the explicit expressions for $L_{\alpha+}$, $L_{\beta-}$, $\Phi_{\text{PPM},j\alpha+}$ and $\Phi_{\text{PPM},j\beta-}$ are directly provided by operating in Eq.~(\ref{eq:corrAB}) the following replacement  
\begin{equation}\label{eq:ABrep}
A(t_r) \to  d(t_r),~~~B(t_s) \to d^\dagger(t_s).
\end{equation}
To exemplify this intuitive argument, we now
consider the case of a single field $\Phi^l(t)$ with $M=1$, and focus on the positive-time correlations $\langle X_n(\tau_2)X_m(\tau_1)\rangle_0$ and $\langle\phi(t)X_n(\tau)\rangle_0$. 

\emph{(i) System dynamics.---}
To analyze the reduced system dynamics, 
we assume the exponential decomposition $\langle X_n(\tau_2)X_m(\tau_1)\rangle_0 = \sum_s w_s \exp{[-(i\Omega_s + \Gamma_s)(\tau_2-\tau_1)]}$, with $\Omega_s\in\mathbb{R},~\Gamma_s\in\mathbb{R}_+,~w_s\in\mathbb{C}$, and $\tau_2\ge\tau_1$. According to Eq.~(\ref{eq:Cppm}), each exponential can be accounted for by one right-purified pseudomode $d_{\alpha+}$ with complex frequency $\Omega_s-i\Gamma_s$ and labeled by the multi-index $\alpha=(n,m,s)$. To determine the generator $L_{\alpha+}$, we can expand~Eq.~(\ref{eq:Ft}) to collect all terms having $\langle X_n(\tau_2)X_m(\tau_1)\rangle_0$ as a prefactor, and then apply the replacement in Eq.~(\ref{eq:ABrep}) to obtain 
\begin{equation}
\begin{array}{l}
\displaystyle\langle X_n(\tau_2)X_m(\tau_1)\rangle_0 S_m(\tau_1)[S_n(\tau_2),\boldsymbol{\cdot}]
\\[2mm]
\displaystyle~~~~~\xlongrightarrow{\text{\normalsize Eq.~(10)}} 
\langle \uwave{d(\tau_2)} \uline{d^\dagger(\tau_1)}\rangle_0 \uline{S_m(\tau_1)}[\uwave{S_n(\tau_2)},\boldsymbol{\cdot}].
\end{array}
\end{equation}
Now, recombining the coupling operators by their time dependence, as marked by different underlines in the second line, directly leads to the generator $L_{\alpha+}[\boldsymbol{\cdot}] = (\Omega_s-i\Gamma_s)d_{\alpha+}^\dagger d_{\alpha+}[\boldsymbol{\cdot}] + \lambda_s' d_{\alpha+} [S_n,\boldsymbol{\cdot}] + \lambda_s'' d_{\alpha+}^\dagger S_m[\boldsymbol{\cdot}]$, 
with $\lambda_s',~\lambda_s''$ the coupling strengths satisfying $w_s=\lambda_s'\lambda_s''$, as promised. 

In the case of one Hermitian system and bath coupling operator $S=S_n,~X=X_n$ with $N=1$, 
Eq.~(\ref{eq:PPMme}) is equivalent to the recently reported free-pole HEOM~\cite{FreePoles}, see the Supplemental Material~\cite{supp_mat} for a formal proof. 
This surprising connection between these models facilitates the optimization of the pseudomode model while also providing a pseudomode perspective to the otherwise abstract auxiliary density operators in the HEOM.

\emph{(ii) Bath input-output.---}
Here, we exemplify the analysis of input-output properties by considering the case of a single field $\Phi^l(t)$. By assuming the corresponding cross-correlation to satisfy 
$\langle\phi(t)X_n(\tau)\rangle_0 = \sum_s w_s \exp{[-(i\Omega_s+\Gamma_s)(t-\tau)]}$, we obtain the right-purified modes $d_{\alpha+}$ with complex frequencies $\Omega_{s}-i\Gamma_s$ and labeled by the multi-index $\alpha=(+,n,s)$. Following similar steps as in the previous analysis, we have
\begin{equation}\label{eq:phiX}
\displaystyle\langle \phi(t)X_n(\tau)\rangle_0 S_n(\tau)[\boldsymbol{\cdot}]
\xlongrightarrow{\text{\normalsize Eq.~(10)}} 
\langle d(t)\uline{d^\dagger(\tau)}\rangle_0 \uline{S_n(\tau)}[\boldsymbol{\cdot}]. 
\end{equation}
Here, on the left-hand side, we collected the terms in Eq.~(\ref{eq:Sat}) associated with $\langle\phi(t)X_n(\tau)\rangle_0$ 
while, on the right-hand side, we used the 
replacement in Eq.~(\ref{eq:ABrep}). As indicated by the underlines, the operators $d^\dagger$ and $S_n$ can now be combined to define the generator $L_{\alpha+}[\boldsymbol{\cdot}] = (\Omega_s-i\Gamma_s)d_{\alpha+}^\dagger d_{\alpha+}[\boldsymbol{\cdot}] + \lambda_s'd_{\alpha+}^\dagger S_n[\boldsymbol{\cdot}]$. The remaining coupling operator in Eq.~(\ref{eq:phiX}), $d$, corresponds to a field superoperator $\Phi_{\text{PPM},\alpha+}^l[\boldsymbol{\cdot}] = \lambda_s''d_{\alpha+}[\boldsymbol{\cdot}]$, with $\lambda_s',~\lambda_s''$ the coupling strengths satisfying $w_s=\lambda_s'\lambda_s''$. 
These are the generators and the pseudomode fields required to reproduce the correct input-output statistics of the original open system for the simulation of the field $\Phi^l(t)$, concluding the construction of the model.

\begin{figure}
\includegraphics[clip,width=8.5cm]{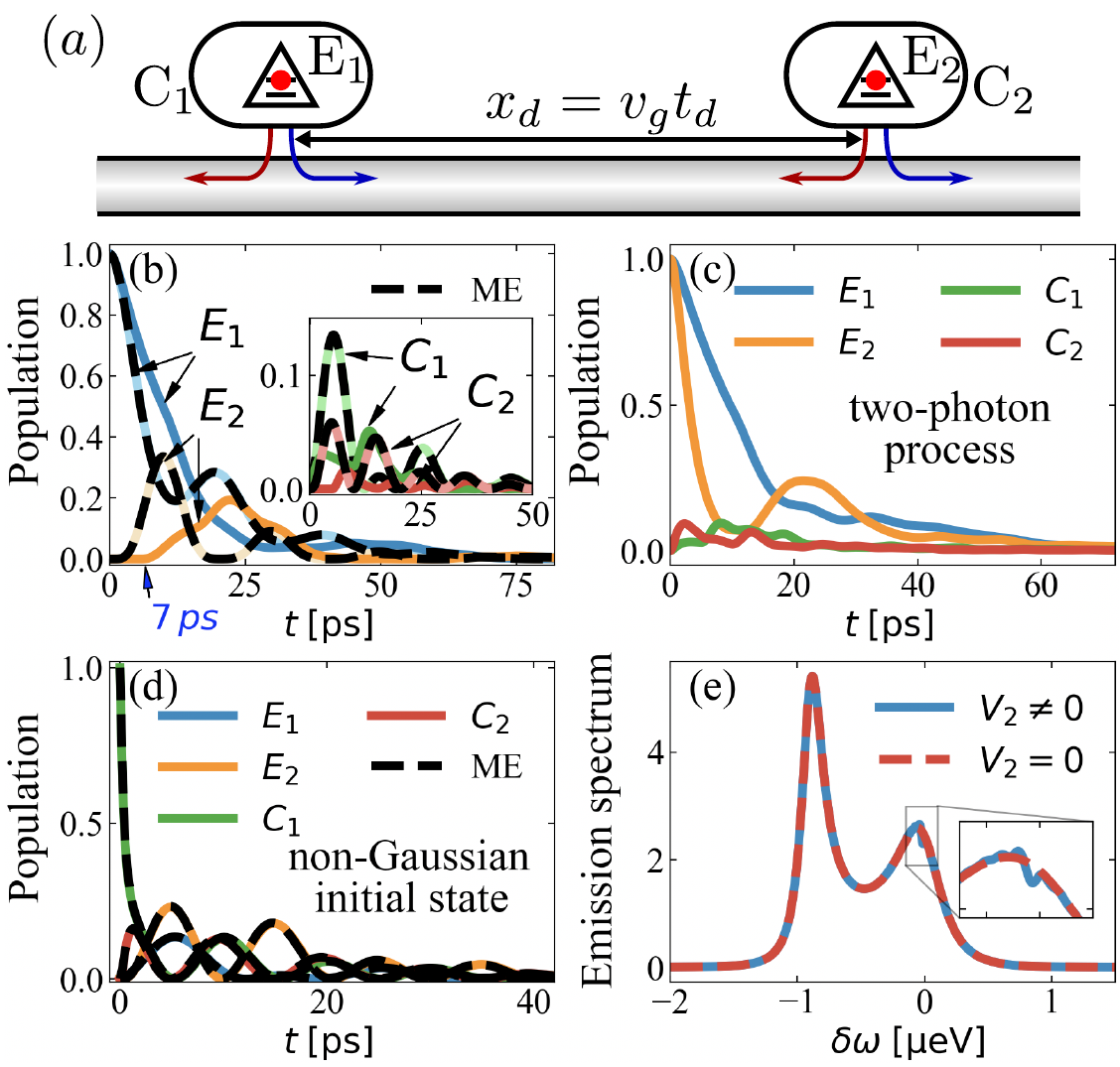}%
\caption{(a) Schematics of the coupled-cavity QED model. (b) Emitter population and cavity occupation (inset) in 
the resonant case for $x_d=4\lambda_0$ (light colors) and $x_d=1500\lambda_0$ (dark colors). Black dashed lines correspond to the results of the Lindblad equation in the Supplemental Material~\cite{supp_mat}. (c) Same as (b) but for both emitters being initially excited and for $x_d=1500\lambda_0$. (d) Same as (b) but for 
a bath initially prepared in the state 
$\rho_\text{B}=c_1^\dagger|0\rangle\langle0| c_1$ and for $x_d=4\lambda_0$. (e) Emission spectrum of $C_2$ in the presence (solid) and absence (dashed) of $E_2$ when $E_1$ is subject to an incoherent driving. Inset: zoom  highlighting the interference pattern. The small oscillation is an artifact caused by the  finite spectral resolution. 
All parameters are consistent with the experimental setting in Ref.~\cite{Yu:21}. 
}\label{fig:results}
\end{figure}

\emph{Numerical implementation.}
We demonstrate the practicality of our approach by simulating the structured cavity-QED
model~\cite{PhysRevLett.98.083603,Yao:09,Yu:21} shown in Fig.~\ref{fig:results}(a). The model consists of two high-quality cavities $C_n$, $n=1,2$, with resonant frequencies $\omega_{c,n}$, side-coupled to a one-dimensional linear waveguide at two locations separated by a distance $x_d$. Each cavity is further locally coupled to a two-level quantum emitter $E_{n}$, with excitation energy $\omega_{e,n}$. We consider a Jaynes-Cummings system-bath interaction 
$H_\text{SB} = \sum_n(\sigma_n^+X_n + \sigma_n^-X_n^\dagger)$ with $X_n=V_nc_n,~V_n\in\mathbb{R}$, where $c_n$ ($c_n^\dagger$) denote the cavity annihilation (creation) operators. By interpreting the cavities and the waveguide as a Gaussian bosonic environment coupled to the two-emitter system, we are going to simulate different initial conditions, namely a non-Gaussian single photon state for the bath and a two-photon state for the system. This model and the parameters used here are compatible with a recent experiment~\cite{Yu:21}, see the Supplemental Material~\cite{supp_mat} for more details.

In Fig.~\ref{fig:results}(b), we plot the emitter population and cavity occupation (inset) dynamics in the resonant case 
$\omega_0 = \omega_{e,n}=\omega_{c,n}$, for small delay $x_d=4\lambda_{0}$ (light colors) and large delay $x_d=1500\lambda_{0}$ (dark colors) with $\lambda_{0}=2\pi v_g/\omega_{0}$ the wavelength and $v_g$ the group velocity. According to Eq.~(\ref{eq:Ecorr}), the cavity occupations can be written in terms of two field superoperators $\Phi_1^l(t) = c_n^\dagger(t)$ and $\Phi_2^l(t)=c_n(t)$. In both cases, the propagating phase $\theta = 2\pi x_d/\lambda_{0} = p\pi$ with $p\in\mathbb{Z}$ implies the existence of a Fano eigenmode with eigenenergy $\omega_{0}$~\cite{Yu:21}. As a consequence, for $x_d=4\lambda_{0}$, we observe  coherent Rabi oscillations between the two emitters, which is consistent with a description in terms of a Lindblad master equation (black dashed) valid for negligible delay-time $\kappa_{c,n}t_d\ll1$~\cite{Yu:21,supp_mat}. Larger $t_d$ leads to a smaller Rabi splitting, so that only one oscillation is visible for $x_d=1500\lambda_{0}$. 
Note that the increase in population for both $E_2$ and $C_2$ at $t\approx\SI{7}{\pico\second}$ is a consequence of the emission from $E_1$, a clear signature of the retardation effect caused by photon propagation. 

In Fig.~\ref{fig:results}(c), we illustrate the two-photon process triggered by exciting both emitters initially for $x_d=1500\lambda_0$, going beyond the scope of Refs.~\cite{Yao:09,PhysRevLett.128.167403}
which focus on one-photon processes or the emitter dynamics only. In Fig.~\ref{fig:results}(d), we demonstrate the possibility to model a non-trivial environmental input by considering the preparation of the bath in a non-Gaussian state 
$\rho_\text{B} = c_1^\dagger|0\rangle\langle0|c_1$ for $x_d=4\lambda_0$. In our formal notation, this corresponds to 
$M=4$ field superoperators $\Phi_1^l(t)=c_n^\dagger(t)$, 
$\Phi_2^l(t)=c_n(t)$, $\Phi_3^l(0)=c_1^\dagger(0)$, and $\Phi_4^r(0)=c_1(0)$. 

In Fig.~\ref{fig:results}(e), we exemplify the possibility to compute bath spectra by showing the steady-state emission spectrum of $C_2$, defined as the Fourier transform of $\langle c_2^\dagger(t+\tau)c_2(t)\rangle_\text{ss}$, when $E_1$ is incoherently driven and off-resonant with $C_1$. Here we consider a scenario in which the two cavities are slightly detuned with respect to each other, while the $E_2$ splitting is approaching the resonance with 
the two nearly degenerate cavity eigenmodes. Interestingly, if $E_2$ is decoupled ($V_2=0$), the emission spectrum shows two resonances, around $\delta\omega = \delta\omega_{e,1},~0$, where 
$\delta\omega=\omega-\omega_r$ and $\delta\omega_{e,1}=\omega_{e,1}-\omega_r$ denote the 
detunings with respect to the reference-frequency $\omega_r=(\omega_{c,1}+\omega_{c,2})/2$. 
On the other hand, a non-zero coupling between $E_2$ and the bath 
($V_2\neq0$) gives rise to a fine interference fringe on top of the Fano resonance, indicating that $E_2$ is excited by the radiation emitted from $E_1$. This fascinating phenomenon, as a strong evidence of the remote coupling mediated by Fano mode in a realistic system, has been observed recently~\cite{Yu:21}.

\emph{Conclusion.}
We have presented a purified version of the pseudomode model for numerically exact simulations of the reduced system dynamics and the bath input-output properties. The illustrated example demonstrates its potential application for non-perturbative treatment of multi-photon transfer processes in coupled nanostructures or quantum networks. In addition, our method also reveals a connection between the pseudomode theory and the HEOM to be used for further optimization of both methods.

\begin{acknowledgments}
M.C. acknowledges support from NSFC (Grant No. 11935012) and NSAF (Grant No. U2330401). N.L. is supported by the RIKEN Incentive Research Program and by MEXT KAKENHI Grants No. JP24H00816 and No. JP24H00820.
\end{acknowledgments}

\bibliography{./refs}

\end{document}


\title{Supplemental Information for \\
``A purified input-output pseudomode model for structured open quantum systems"}

\author{Pengfei Liang}
\email{pfliang@imu.edu.cn}
\affiliation{Center for Quantum Physics and Technologies, School of Physical Science and Technology, Inner Mongolia University, Hohhot 010021, China}
\affiliation{Graduate School of China Academy of Engineering Physics, Haidian District, Beijing, 100193, China}

\author{Neill Lambert}
\email{nwlambert@gmail.com}
\affiliation{Theoretical Physics Laboratory, Cluster for Pioneering Research, RIKEN, Wakoshi, Saitama 351-0198, Japan}
\affiliation{Quantum Computing Center, RIKEN, Wakoshi, Saitama, 351-0198, Japan}

\author{Si Luo}
\affiliation{Department of Chemistry, Fudan University, Shanghai 200433, P. R. China}

\author{Lingzhen Guo}
\affiliation{Center for Joint Quantum Studies and Department of Physics, School of Science, Tianjin University, Tianjin 300072, China}

\author{Mauro Cirio}
\email{cirio.mauro@gmail.com}
\affiliation{Graduate School of China Academy of Engineering Physics, Haidian District, Beijing, 100193, China}

\date{\today}
\maketitle

In this supplemental material, we formally complement the derivations presented in the main text to include explicit technical details. Specifically, in Sec.~\ref{sec:PMadiabaticlimits} we describe the procedure to achieve the purification of zero-temperature pseudomodes, and use it to show the equivalence of Eq.~(5) and Eq.~(8) in the main text. In Sec.~\ref{sec:complexity} we analyze the computational complexity of the conventional pseudomode model and our purified pseudomode model in the context of tensor-network simulations. In Sec.~\ref{sec:rigorousconstruction}, we explicitly justify the validity of the replacement rules in Eq.~(11) in the main text. In Sec.~\ref{sec:relationtofpHEOM}, we show that our purified pseudomode model and the free-pole HEOM reported in Ref.~[46] yield identical reduced system dynamics for the system-bath interaction Hamiltonian $H_\text{SB}=SX$. 
In Sec.~\ref{sec:coupledcavityQED}, we finish by reviewing the coupled-cavity waveguide model considered in the main text to also include the explicit  calculation of the corresponding bath correlations, and by further  summarizing, the model parameters used in the numerical simulations.

\section{Purification of Zero-Temperature Pseudomodes}\label{sec:PMadiabaticlimits}

Here, we describe the details of the purification of zero-temperature pseudomodes. We do this by considering the analytical continuation of their frequency and decay rate along the paths shown in Fig.~2 in the main text. This continuation allows to proceed with a partial adiabatic elimination of the pseudomode degrees of freedom in the Liouville space. We further show that this procedure also explicitly justifies the equivalence given in Eq.~(5) and Eq.~(8) in the main text.

We start by considering the analytical continuation of the frequency and decay rate of the pseudomodes along the path $\Omega\to \Omega\pm ia$ and $\Gamma\to \Gamma+a$, respectively. This ultimately affects the Lindblad master equation which becomes
\begin{equation}\label{eq:ME}
\begin{array}{lll}
\displaystyle \frac{d\rho}{dt} &\displaystyle = -i[H_\text{S} + (\Omega_++ia) d_+^\dagger d_+ + (\Omega_--ia) d_-^\dagger d_-, \rho] \\
&\displaystyle~~~ -i\big( \lambda_{1+}S_{1+}d_+\rho + \lambda_{2+}S_{2+}d_+^\dagger\rho - \lambda_{3+}\rho S_{3+}d_+ - \lambda_{4+}\rho S_{4+}d_+^\dagger \big) \\
&\displaystyle~~~ -i\big( \lambda_{1-}S_{1-}d_-\rho + \lambda_{2-}S_{2-}d_-^\dagger\rho - \lambda_{3-}\rho S_{3-}d_- - \lambda_{4-}\rho S_{4-}d_-^\dagger \big) \\
&\displaystyle~~~ + (\Gamma_++a)(2d_+\rho d_+^\dagger - d_+^\dagger d_+\rho - \rho d_+^\dagger d_+) + (\Gamma_-+a)(2d_-\rho d_-^\dagger - d_-^\dagger d_-\rho - \rho d_-^\dagger d_-).
\end{array}
\end{equation}
Here, we have considered a  general interaction Hamiltonian in terms of the system coupling operators $S_{j\pm}$ with coupling strengths $\lambda_{j\pm}$. Importantly, $d_\pm$ are initially prepared in their vacuum states, i.e., $\rho(0)=\rho_\text{S}\otimes|0_+0_-\rangle\langle0_+0_-|$
in terms of the system initial state $\rho_\text{S}$.

To proceed, we first write this equation in a vectorized form which corresponds to writing
Eq.~(\ref{eq:ME}) in the Liouville space $\mathcal{L} = \mathcal{H}_\text{S}\otimes\mathcal{H}_\text{S}\otimes\mathcal{H}_{\text{PM}+}\otimes\mathcal{H}_{\text{PM}+}\otimes\mathcal{H}_{\text{PM}-}\otimes\mathcal{H}_{\text{PM}-}$, where $\mathcal{H}_\text{S}$ and $\mathcal{H}_{\text{PM}\pm}$ denote the Hilbert space of the system and the two pseudomodes, respectively. To do this, we map the left (right) action of 
an operator $O$, with support on either $\mathcal{H}_\text{S}$ or $\mathcal{H}_{\text{PM}\pm}$,  
to $\bar{O}\equiv O\otimes I$ ($\tilde{O}^\intercal \equiv I\otimes O^\intercal$), which act on either $\mathcal{H}_\text{S}\otimes\mathcal{H}_\text{S}$ or $\mathcal{H}_{\text{PM}\pm}\otimes\mathcal{H}_{\text{PM}\pm}$. Here, $I$ denotes the corresponding identity operator and $O^\intercal$ the transpose of $O$. On the other hand, the density matrix $\rho$ is vectorized as 
\begin{equation}
\begin{array}{lll}
&\displaystyle \rho = \sum_{s,s',n_\pm,n_\pm'} \rho_{sn_+n_-,s'n_+'n_-'} |s\rangle\langle s'|\otimes|n_+\rangle\langle n_+'|\otimes|n_-\rangle\langle n_-'| \\
&\displaystyle~~~~~~~~~\to |\rho\rangle\rangle =  \sum_{s,s',n_\pm,n_\pm'} \rho_{sn_+n_-,s'n_+'n_-'} |ss'\rangle\rangle\otimes|n_+n_+'\rangle\rangle\otimes|n_-n_-'\rangle\rangle,  
\end{array}
\end{equation}
with $\rho_{sn_+n_-,s'n_+'n_-'}$ the matrix elements of $\rho$ with respect to the basis states $|sn_+n_-\rangle$ and $|s'n_+'n_-'\rangle$. In this alternative language, Eq.~(\ref{eq:ME}) can be equivalently written as
\begin{equation}
\begin{array}{lll}
\displaystyle i\frac{d|\rho\rangle\rangle}{dt} = \mathfrak{L}|\rho\rangle\rangle = (\mathfrak{L}_\text{S}+\mathfrak{L}_{\text{PM}+}^a+\mathfrak{L}_{\text{PM}-}^a)|\rho\rangle\rangle, 
\end{array}
\end{equation}
where $\mathfrak{L}_\text{S} = \bar{H}_\text{S} - \tilde{H}_\text{S}^\intercal$ and
\begin{equation}
\begin{array}{lll}
\displaystyle \mathfrak{L}_{\text{PM}+}^a &\displaystyle = (\Omega_+-i\Gamma_+)\bar{d}_+^\dagger\bar{d}_+ - (\Omega_++i(\Gamma_++2a))\tilde{d}_+^\dagger\tilde{d}_+ + 2i(\Gamma_++a)\bar{d}_+\tilde{d}_+  \\
&\displaystyle~~~ + \lambda_{1+}\bar{S}_{1+}\bar{d}_{+} + \lambda_{2+}\bar{S}_{2+}\bar{d}_{+}^\dagger - \lambda_{3+}\tilde{S}_{3+}^\intercal\tilde{d}_{+}^\dagger - \lambda_{4+}\tilde{S}_{4+}^\intercal \tilde{d}_{+},  \\
\displaystyle \mathfrak{L}_{\text{PM}-}^a &\displaystyle = (\Omega_--i(\Gamma_-+2a))\bar{d}_-^\dagger\bar{d}_- - (\Omega_-+i\Gamma_-)\tilde{d}_-^\dagger\tilde{d}_- + 2i(\Gamma_-+a)\bar{d}_-\tilde{d}_-  \\
&\displaystyle~~~ + \lambda_{1-}\bar{S}_{1-}\bar{d}_{-} + \lambda_{2-}\bar{S}_{2-}\bar{d}_{-}^\dagger - \lambda_{3-}\tilde{S}_{3-}^\intercal\tilde{d}_{-}^\dagger - \lambda_{4-}\tilde{S}_{4-}^\intercal \tilde{d}_{-}, 
\end{array}
\end{equation}
are the system and pseudomode Liouvillian operators, respectively. The initial state  can now be written as the vector
$|\rho(0)\rangle\rangle = |\rho_\text{S}\rangle\rangle\otimes|\bar0_+\tilde0_+\rangle\rangle\otimes|\bar0_-\tilde0_-\rangle\rangle$ where, for later convenience, we used different notations to denote the vacuum of $\bar{d}_\pm$ and $\tilde{d}_\pm$. 

To proceed, we further split the Liouvillian operator $\mathfrak{L}$ as $\mathfrak{L} = \mathfrak{L}_\text{f}+\mathfrak{L}_0+\mathfrak{V}$, where
\begin{equation}\label{eq:Lf0V}
\begin{array}{lll}
\mathfrak{L}_\text{f} &\displaystyle = \bar{H}_\text{S} - \tilde{H}_\text{S}^\intercal + (\Omega_+-i\Gamma_+)\bar{d}_+^\dagger\bar{d}_+ - (\Omega_-+i\Gamma_-)\tilde{d}_-^\dagger\tilde{d}_- + \lambda_{1+}\bar{S}_{1+}\bar{d}_{+} + \lambda_{2+}\bar{S}_{2+}\bar{d}_{+}^\dagger - \lambda_{3-}\tilde{S}_{3-}^\intercal\tilde{d}_{-}^\dagger - \lambda_{4-}\tilde{S}_{4-}^\intercal \tilde{d}_{-}, \\
\mathfrak{L}_0 &\displaystyle = - (\Omega_++i(\Gamma_++2a))\tilde{d}_+^\dagger\tilde{d}_+ + (\Omega_--i(\Gamma_-+2a))\bar{d}_-^\dagger\bar{d}_- + 2i(\Gamma_++a)\bar{d}_+\tilde{d}_+ + 2i(\Gamma_-+a)\bar{d}_-\tilde{d}_- \\
&\displaystyle~~~  - \lambda_{4+}\tilde{S}_{4+}^\intercal \tilde{d}_+ + \lambda_{1-}\bar{S}_{1-}\bar{d}_-, \\
\mathfrak{V} &\displaystyle = -\lambda_{3+}\tilde{S}_{3+}^\intercal\tilde{d}_+^\dagger + \lambda_{2-}\bar{S}_{2-}\bar{d}_-^\dagger. 
\end{array}
\end{equation}
We do this to ensure that $\mathfrak{L}_\text{f}+\mathfrak{L}_0$ does not contain the creation operators $\tilde{d}_+^\dagger$ and $\bar{d}_-^\dagger$ of the two modes with an enhanced decay rate $\Gamma_++2a$ or $\Gamma_-+2a$, whose degrees of freedom we wish to eliminate. In fact, this observation on their effective decay rate implies that 
these two modes will be frozen in their initial vacuum state $|\tilde{0}_+\bar{0}_-\rangle\rangle$ under the dynamics generated by $\mathfrak{L}_\text{f}+\mathfrak{L}_0$. In contrast, $\mathfrak{V}$ triggers transitions out of their vacuum state. Our goal is thus to eliminate the term $\mathfrak{V}$ in the limit $a\to+\infty$ such that the modes $\tilde{d}_+,~\bar{d}_-$ are dynamically decoupled from the rest degrees of freedom. 

To this end, we look for a Liouvillian operator $\mathfrak{S}$, which, crucially, we require to be $\mathcal{O}(a^{-1})$, to define the Schrieffer-Wolff (SW) transformation 
\begin{equation}\label{eq:SW}
e^{\mathfrak{S}}\mathfrak{L}e^{-\mathfrak{S}} = \mathfrak{L} + [\mathfrak{S},\mathfrak{L}] + \frac{1}{2!}[\mathfrak{S},[\mathfrak{S},\mathfrak{L}]] + \cdots, 
\end{equation}
such that the following operator equation is satisfied
\begin{equation}\label{eq:Seq}
[\mathfrak{S},\mathfrak{L}_\text{f}+\mathfrak{L}_0] = -\mathfrak{V} + \mathfrak{V}'+\mathcal{O}(a^{-1}). 
\end{equation}
Here, the operator $\mathfrak{V}'$ represents the emerging mediated interactions, and for consistency we require, not to contain neither $\tilde{d}_+^\dagger$ or $\bar{d}_-^\dagger$, and to be $\mathcal{O}(a^0)$. The reason for requiring the existence of the commutator in Eq.~(\ref{eq:Seq}) is that it ensures nested commutators appearing in the SW transformation, such as  $[\mathfrak{S},\cdots,[\mathfrak{S},\mathfrak{L}_\text{f}+\mathfrak{L}_0]]$ with at least two $\mathfrak{S}$, or $[\mathfrak{S},\cdots,[\mathfrak{S},\mathfrak{V}]]$ with at least one $\mathfrak{S}$, to be at least of order $\mathcal{O}(a^{-1})$. Hence, in the limit $a\to+\infty$, they vanish in the expansion of the SW transformation in Eq.~(\ref{eq:SW}), leading to the simple equation 
\begin{equation}
e^{\mathfrak{S}}\mathfrak{L}e^{-\mathfrak{S}}=\mathfrak{L}_\text{f}+\mathfrak{L}_0+\mathfrak{V}', 
\end{equation}
so that the modes $\tilde{d}_+,~\bar{d}_-$ are decoupled from the rest degrees of freedom. Ultimately, this allows to project $e^{\mathfrak{S}}\mathfrak{L}e^{-\mathfrak{S}}$ onto $|\tilde{0}_+\bar{0}_-\rangle$ to arrive at the new dynamical semigroup generator 
\begin{equation}\label{eq:newGen}
\mathfrak{L}_\text{eff} = \langle\tilde{0}_+\bar{0}_-|e^{\mathfrak{S}}\mathfrak{L}e^{-\mathfrak{S}}|\tilde{0}_+\bar{0}_-\rangle = \mathfrak{L}_\text{f}+\mathfrak{V}'. 
\end{equation}
The remaining task is to solve Eq.~(\ref{eq:Seq}). We have found the solution
\begin{equation}
\mathfrak{S} = \frac{-1}{2ia}\mathfrak{V} = \frac{-1}{2ia} \Big( -\lambda_{3+}\tilde{S}_{3+}^\intercal\tilde{d}_+^\dagger + \lambda_{2-} \bar{S}_{2-}\bar{d}_-^\dagger \Big), 
\end{equation}
which satisfies $[\mathfrak{S},\mathfrak{L}_\text{f}] = 0$ and 
\begin{equation}
\begin{array}{lll}
&\displaystyle \lim_{a\to+\infty}[\mathfrak{S},\mathfrak{L}_0] = [-\lambda_{3+}\tilde{S}_{3+}^\intercal\tilde{d}_+^\dagger + \lambda_{2-} \bar{S}_{2-}\bar{d}_-^\dagger, \tilde{d}_+^\dagger\tilde{d}_++\bar{d}_-^\dagger \bar{d}_- - \bar{d}_+\tilde{d}_+ - \bar{d}_-\tilde{d}_-] \\
&\displaystyle~~~ = \lambda_{3+} \tilde{S}_{3+}^\intercal \tilde{d}_+^\dagger - \lambda_{2-} \bar{S}_{2-}\bar{d}_-^\dagger
- \lambda_{3+} \tilde{S}_{3+}^\intercal \bar{d}_+ + \lambda_{2-} \bar{S}_{2-} \tilde{d}_- = -\mathfrak{V} + \mathfrak{V}', 
\end{array}
\end{equation}
with $\mathfrak{V}' = - \lambda_{3+} \tilde{S}_{3+}^\intercal \bar{d}_+ + \lambda_{2-} \bar{S}_{2-} \tilde{d}_-$. Note that, since $\mathfrak{S}$ is of order $\mathcal{O}(a^{-1})$, we have $\lim_{a\to+\infty} e^{\mathfrak{S}}|\rho\rangle\rangle = |\rho\rangle\rangle$, meaning that in the limit $a\to+\infty$ the vectorized density matrix remains invariant under the SW transformation. 

By combining Eq.~(\ref{eq:Lf0V}) and Eq.~(\ref{eq:newGen}), the equation $id|\rho\rangle\rangle/dt = \mathfrak{L}_\text{eff}|\rho\rangle\rangle$ can be rewritten in the original density matrix representation as 
\begin{equation} 
i\frac{d\rho}{dt} = [H_\text{S},\rho] + L_{\text{g},+}\rho + L_{\text{g},-}\rho, 
\end{equation}
where the two superoperators read
\begin{equation}\label{eq:Lgpm}
\begin{array}{lll}
\displaystyle L_{\text{g},+}[\boldsymbol{\cdot}] &\displaystyle= (\Omega_+ - i\Gamma_+)d_+^\dagger d_+[\boldsymbol{\cdot}] + \lambda_{1+}d_+S_{1+}[\boldsymbol{\cdot}] + \lambda_{2+}d_+^\dagger S_{2+}[\boldsymbol{\cdot}] - \lambda_{3+}d_+[\boldsymbol{\cdot}]S_{3+}, \\
\displaystyle L_{\text{g},-}[\boldsymbol{\cdot}] &\displaystyle= -(\Omega_-+i\Gamma_-)[\boldsymbol{\cdot}]d_-^\dagger d_- - \lambda_{3-}[\boldsymbol{\cdot}]S_{3-}d_- - \lambda_{4-}[\boldsymbol{\cdot}]S_{4-}d_-^\dagger + \lambda_{2-}S_{2-}[\boldsymbol{\cdot}]d_-^\dagger. 
\end{array}
\end{equation}
The expressions in Eq.~(\ref{eq:Lgpm}) constitute the main results of this section. 

We now prove the equivalence of Eq.~(5) and Eq.~(8) in the main text, in the sense that they yield identical reduce system dynamics. The discussion here is a simplified version of the general one presented in Sec.~\ref{sec:rigorousconstruction}. First, we note that the bath correlation $\mathcal{C}(t) = \lambda^2\exp{[-i\Omega t - \Gamma\lvert t\rvert]}$ can be approximated as
\begin{equation}\label{eq:Capprox}
C(t) =  \lambda^2 e^{-i(\Omega+ia)t - (\Gamma+a)\lvert t\rvert} + \lambda^2 e^{-i(\Omega-ia)t - (\Gamma+a)\lvert t\rvert} + \mathcal{O}(e^{-a\lvert t\rvert}),
\end{equation}
with $a$ some large positive real number. In parallel to the discussion of the single pseudomode model in the main text, this approximation implies that up to an error of order $\mathcal{O}(\exp{[-a\lvert t\rvert]})$, the reduced density matrix $\rho_\text{S}(t)$ of the original open quantum system characterized by the bath correlation $\mathcal{C}(t)$ can be reproduced by a two-pseudomode model, described by the Lindblad master equation
\begin{equation}\label{eq:twoPMmodel}
i\frac{d\rho_\text{eff}^a}{dt} = [H_\text{S},\rho_\text{eff}^a] + L_\text{PM,+}^a\rho_\text{eff}^a + L_\text{PM,-}^a\rho_\text{eff}^a, 
\end{equation}
with $\rho_\text{S} = \text{Tr}_\pm[\rho_\text{eff}^a] + \mathcal{O}(\exp{[-a\lvert t\rvert]})$ and the generators $L_{\text{g},\pm}[\boldsymbol{\cdot}] = [(\Omega\pm ia)d_\pm^\dagger d_\pm,\boldsymbol{\cdot}] + i(\Gamma+a)(2d_\pm[\boldsymbol{\cdot}]d_\pm^\dagger - d_\pm^\dagger d_\pm[\boldsymbol{\cdot}] - [\boldsymbol{\cdot}]d_\pm^\dagger d_\pm)$. Note that the trace $\text{Tr}_\pm[\boldsymbol{\cdot}]$ is taken over both $d_+$ and $d_-$. 

Consistently, Eq.~(\ref{eq:twoPMmodel}) is exactly the same as Eq.~(\ref{eq:ME}) if we define $\Omega_\pm=\Omega$, $\Gamma_\pm=\Gamma$, $\lambda_{j\pm}=\lambda$ and $S_{j\pm}=S$. By taking the limit $a\to+\infty$, the approximation in Eq.~(\ref{eq:Capprox}) becomes exact and the purification procedure described above can be applied to yield Eq.~(8) in the main text, i.e., $\rho_\text{eff} = \lim_{a\to+\infty} \rho_\text{eff}^a$ and $L_\pm = \lim_{a\to+\infty} L_{\text{PM},\pm}^a$.

\section{Computational Complexity Analysis}\label{sec:complexity}
\begin{figure}
\includegraphics[clip,width=8.5cm]{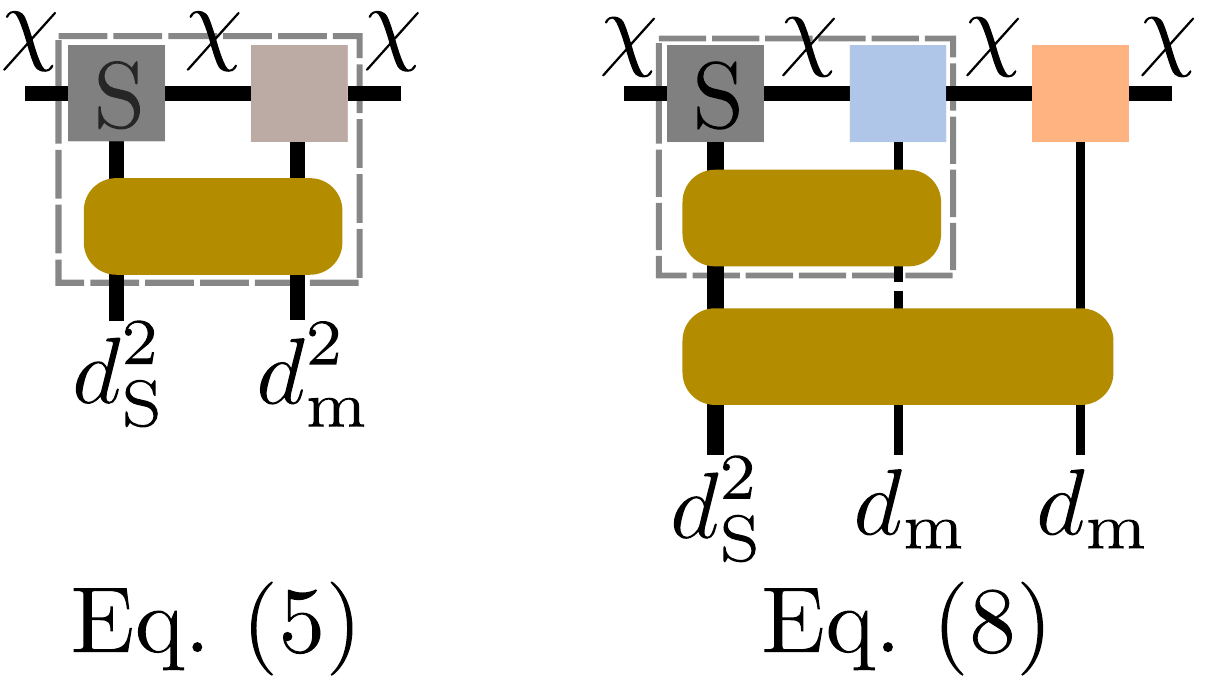}%
\caption{Diagrams for the tensor-network simulation of Eq.~(5) and Eq.~(8) in the main text using the tDMRG algorithm. 
}\label{fig:complexity}
\end{figure}

Here, we compare the computational complexity of the two models defined in Eqs.~(5) and (8) in the main text
from the perspective of tensor-network algorithms. In particular, we focus on the time-dependent density matrix renormalization group (tDMRG) algorithm, which has been adopted in our numerical simulations to produce Fig.~3 in the main text. To make use of the tDMRG, we map the density matrix in Eqs.~(5) and (8) to a vector in the Liouville space so as to represent it as a matrix product state (MPS), as shown in Fig.~\ref{fig:complexity}. For the pseudomode model given in Eq.~(5) [left diagram], the local Hilbert dimension of the system and the pseudomode are $d_\text{S}^2$ and $d_\text{m}^2$, respectively, where $d_\text{S}$ and $d_\text{m}$ denote the truncation of the system Hilbert space and the pseudomode Fock space. In contrast, for the purified model in Eq.~(8) [right diagram], the local Hilbert dimension of the two purified modes are $d_\text{S}$ as a consequence of the purification.   

To evolve the MPS, the propagators in Eqs.~(5) and (8) are Trotterized and successively contracted with the MPS. Fig.~\ref{fig:complexity} also illustrates the contraction of two-site gates with the MPS, which yield two-site plain tensors having dimension $d_\text{S}^2\chi\times d_\text{m}^2\chi$ [Eq.~(5)] or $d_\text{S}^2\chi\times d_\text{m}\chi$ [Eq.~(8)] with $\chi$ the bond dimension, as marked by the gray boxes. To restore the structure of the MPS, singular value decompositions (SVDs) of such plain tensors are required, which are the most expensive operations in the tDMRG algorithm. Specifically, the computational complexity of such SVDs is $O\!\left(\min \!\left[d_\text{S}^4d_\text{PM}^2\chi^3,d_\text{S}^2d_\text{PM}^4\chi^3\right]\right)$ in Eq.~(5) and $O\!\left(\min \!\left[2d_\text{S}^4d_\text{PM}\chi^3,2d_\text{S}^2d_\text{PM}^2\chi^3\right]\right)$ in Eq.~(8), where the factor $2$ in the latter arises since two purified modes are used in Eq.~(8). In many situations, especially in some difficult parameter regimes, a much larger truncation of the mode Fock space is required to achieve convergence, i.e., $d_\text{m}>d_\text{S}$. 
Hence, the reduction of the scaling exponents of $d_\text{m}$ by $1$ or $2$ in $O\!\left(\min \!\left[2d_\text{S}^4d_\text{PM}\chi^3,2d_\text{S}^2d_\text{PM}^2\chi^3\right]\right)$, compared to $O\!\left(\min\! \left[d_\text{S}^4d_\text{PM}^2\chi^3,d_\text{S}^2d_\text{PM}^4\chi^3\right]\right)$, indicates that the purified pseudomode model in Eq.~(8) has better numerical performance in practical simulations. 

The above analysis can be generalized to more complex bath correlations, which require more exponentials to represent. However, since the analysis here can be applied to each exponential, we can still conclude that the corresponding purified pseudomode model should have a better numerical performance.

\section{Explicit Construction of Purified Pseudomode Models}\label{sec:rigorousconstruction}
In this section, we provide formal and explicit details for the validity of the replacement rules whose practical derivation is  provided in the main text. 

In order to provide intuition on the technical construction, we start by observing that the
purified pseudomode correlations $\mathcal{C}_\text{PPM}^\pm(t) = \lambda^2 \exp(-i\Omega t-\Gamma\lvert t\rvert)\Theta(\pm t)$ [Eq.~(7) in the main text] can be approximated as 
\begin{equation}\label{eq:PMa}
\mathcal{C}_\text{PM}^\pm(t)\approx\mathcal{C}_{\text{PM},a}^\pm(t) \equiv \lambda^2 e^{-i(\Omega\pm ia)t - (\Gamma+a)\lvert t\rvert}, 
\end{equation}
with an error $\lvert \mathcal{C}_{\text{PM},a}^\pm(t) - \mathcal{C}_{\text{PPM}}^\pm(t)\rvert \sim \mathcal{O}(\exp{[-a\lvert t\rvert]})$. Importantly, $\mathcal{C}_{\text{PM},a}^\pm(t)$ can be viewed as the the free correlation of a zero-temperature pseudomode with complex frequency $\Omega\pm ia$ and decay rate $\Gamma+a$. For convenience, we will call them large-$a$ pseudomodes. Based on this observation, we can define an exact purified pseudomode model through the following two steps: First, we show the (approximate) equivalence of the original open model and a large-$a$ pseudomode model by matching all of their correlations and cross-correlations, which is equivalent to show the existence of a mapping
\begin{equation}
X_n(\tau) \to X_{\text{PM},n}(\tau),~~~\Phi_j^{\alpha_j}(t_j) \to \Phi_{\text{PM},j}^{\alpha_j}(t_j).  
\end{equation} 
Second, once this mapping is found, the purification technique developed in Sec.~\ref{sec:PMadiabaticlimits} can be employed to eliminate the dependence on the free parameter $a$, thereby obtaining the desired purified pseudomode model.

In general, a large-$a$ pseudomode model is described by the Lindblad master equation,
\begin{equation}
i\frac{d\rho_\text{eff}^a}{dt} = [H_\text{S},\rho_\text{eff}^a] + \sum_{\alpha}L_{\text{PM},\alpha+}^a\rho_\text{eff}^a + \sum_{\beta}L_{\text{PM},\beta-}^a\rho_\text{eff}^a, 
\end{equation}
where the superoperators $L_{\text{PM},\alpha+}^a$ and $L_{\text{PM},\beta-}^a$ are defined as
\begin{equation}
\begin{array}{lll}
L_{\text{PM},\alpha+}^a[\boldsymbol{\cdot}] &\displaystyle = \big[(\Omega_\alpha+ia) d_{\alpha+}^\dagger d_{\alpha+} + H_{\text{int},\alpha+},\boldsymbol{\cdot}\big] + i(\Gamma_\alpha+a)(d_{\alpha+}[\boldsymbol{\cdot}]d_{\alpha+}^\dagger - d_{\alpha+}^\dagger d_{\alpha+}[\boldsymbol{\cdot}] - [\boldsymbol{\cdot}]d_{\alpha+}^\dagger d_{\alpha+}), \\
L_{\text{PM},\beta-}^a[\boldsymbol{\cdot}] &\displaystyle = \big[(\Omega_\beta-ia) d_{\beta-}^\dagger d_{\beta-} + H_{\text{int},\beta-},\boldsymbol{\cdot}\big] + i(\Gamma_\beta+a)(d_{\beta-}[\boldsymbol{\cdot}]d_{\beta-}^\dagger - d_{\beta-}^\dagger d_{\beta-}[\boldsymbol{\cdot}] - [\boldsymbol{\cdot}]d_{\beta-}^\dagger d_{\beta-}),  
\end{array}
\end{equation}
in terms of the pseudomode interaction Hamiltonians $H_{\text{int},\alpha+}$ and $H_{\text{int},\beta-}$. Accordingly, the effective field superoperator $\Phi_{\text{PM},j}^{\alpha_j}$ can be written as the sum of its components on each pseudomode Hilbert space, i.e.,
\begin{equation}
\Phi_{\text{PM},j}^{\alpha_j}  = \sum_\alpha \Phi_{\text{PM},j\alpha+}^{\alpha_j} + \sum_\beta \Phi_{\text{PM},j\beta-}^{\alpha_j}.     
\end{equation} 
Here, the Hamilonians $H_{\text{int},\alpha+}$, $H_{\text{int},\beta-}$, and the superoperators $\Phi_{\text{PM},j\alpha+}^{\alpha_j}$, and $\Phi_{\text{PM},j\beta-}^{\alpha_j}$ are unknowns to be determined from the expression of the bath correlations in Eq.~(4) in the main text. As long as the interaction Hamiltonians $H_{\text{int},\alpha+}$ and $H_{\text{int},\beta-}$ are found, $X_{\text{PM},n}$ can be defined by solving $\sum_{\alpha}'H_{\text{int},\alpha+} + \sum_\beta' H_{\text{int},\beta-} = S_nX_{\text{PM},n}$ where the prime means that the sum runs over those interaction Hamiltonian containing the coupling operator $S_n$.

For clarity, in the following, 
we will discuss the reduced system dynamics and the bath input-output separately, and show explicitly that the rigorous construction defined here yields the same purified pseudomode model as the more practical argument provided in the main text.

\subsection{System dynamics}
\begin{table}[t]
\centering
\begin{tabular}{>{\centering}m{0.08\textwidth}>{\centering}m{0.32\textwidth}>{\centering\arraybackslash}m{0.32\textwidth}}
\toprule
\textbf{corr.} & $\langle X_n(\tau_2)X_m(\tau_1)\rangle_0$  & $\langle X_m^\dagger(\tau_1)X_n^\dagger(\tau_2)\rangle_0$   \\
\midrule
\textbf{decomp.} & $\sum_s w_s e^{-(i\Omega_s+\Gamma_s)(\tau_2-\tau_1)}$ & $\sum_s w_s^* e^{-(i\Omega_s-\Gamma_s)(\tau_1-\tau_2)}$  \\
\midrule
\textbf{approx.} & $\sum_s w_s e^{-i(\Omega_s+ia)(\tau_2-\tau_1)-(\Gamma_s+a)\lvert \tau_2-\tau_1\rvert}$ & $\sum_s w_s^* e^{-i(\Omega_s-ia)(\tau_1-\tau_2)-(\Gamma_s+a)\lvert\tau_1-\tau_2\rvert}$  \\
\midrule
\textbf{freq.} & $\Omega_s+ia$ & $\Omega_s-ia$    \\
\midrule
\textbf{decay} & $\Gamma_s+a$ & $\Gamma_s+a$ \\
\midrule
$\mathbf{H_{\text{int}}}$ & $\lambda_s' S_nd_{\alpha+} +\lambda_s''S_md_{\alpha+}^\dagger$ & $\lambda_s^{\prime\prime*}S_m^\dagger d_{\beta-} + \lambda_s^{\prime*} S_n^\dagger d_{\beta-}^\dagger $   \\
\midrule 
\textbf{purif.} & $\lambda_s'd_{\alpha+}[S_n,\boldsymbol{\cdot}] + \lambda_s'' d_{\alpha+}^\dagger S_m[\boldsymbol{\cdot}]$ & $\lambda_s^{\prime*}[S_n^\dagger,\boldsymbol{\cdot}]d_{\beta-}^\dagger - \lambda_s^{\prime\prime*}[\boldsymbol{\cdot}]S_m^\dagger d_{\beta-}$  \\
\bottomrule
\end{tabular}
\caption{Correlations required for reproducing the reduced system dynamics are listed in the first row. Their exponential decompositions and approximations required to define large-$a$ pseudomodes are shown in the second and third rows, respectively. The frequency, decay rate and interaction Hamiltonian of the large-$a$ pseudomodes are given in the fourth to sixth rows. The multi-indices $\alpha$ and $\beta$ are defined, respectively, as $\alpha = (n,m,s)$ and $\beta=(m,n,s)$. In the interaction Hamiltonian, the constraint $w_s=\lambda_s'\lambda_s''$ should be imposed. The last row corresponds to the purification of the commutators $[H_\text{int},\boldsymbol{\cdot}]$. 
}\label{tab:sysdyn}
\end{table}
In the main text, we have assumed the exponential decomposition ($\tau_2\ge\tau_1$)
\begin{equation}\label{eq:XnXmpos}
\langle X_n(\tau_2)X_m(\tau_1)\rangle_0 = \sum_s w_s e^{-(i\Omega_s + \Gamma_s)(\tau_2-\tau_1)},
\end{equation}
for each positive-time correlation, with $\Omega_\alpha\in\mathbb{R},~\Gamma_\alpha\in\mathbb{R}_+,~w_\alpha\in\mathbb{C}$. In order to define large-$a$ pseudomodes to model this correlation, we use the following approximation 
\begin{equation}
\sum_s w_s e^{-(i\Omega_s + \Gamma_s)(\tau_2-\tau_1)}\Theta(\tau_2-\tau_1) \approx \sum_s w_s e^{-i(\Omega_s+ia)(\tau_2-\tau_1) - (\Gamma_s+a)\lvert \tau_2-\tau_1\rvert}. 
\end{equation}
Compared to the definition of $\mathcal{C}_{\text{PM},a}^+(t)$, we conclude that each exponential on the right side corresponds to a large-$a$ pseudomode $d_{\alpha+}$, labeled by the multi-index $\alpha=(n,m,s)$, characterized by the frequency $\Omega_s+ia$, the decay rate $\Gamma_s+a$ and the interaction Hamiltonian $H_{\text{int},\alpha+} = \lambda_s'S_nd_{\alpha+}+\lambda_s''S_md_{\alpha+}^\dagger$, where the coupling strengths satisfy $w_s=\lambda_s'\lambda_s''$. 

Thanks to  the time-reversal property $\langle X_m^\dagger(\tau_1)X_n^\dagger(\tau_2)\rangle_0 = \langle X_n(\tau_2)X_m(\tau_1)\rangle_0^*$, the decomposition in Eq.~(\ref{eq:XnXmpos}) also yields the following decomposition
\begin{equation}
\langle X_m^\dagger(\tau_1)X_n^\dagger(\tau_2)\rangle_0 = \sum_s w_s^* e^{-(i\Omega_s - \Gamma_s)(\tau_1-\tau_2)}, 
\end{equation}
for negative-time correlations. Using the approximation
\begin{equation}
\sum_s w_s^* e^{-(i\Omega_\alpha - \Gamma_s)(\tau_1-\tau_2)}\Theta(\tau_2-\tau_1) \approx \sum_\alpha w_s^* e^{-i(\Omega_s-ia)(\tau_1-\tau_2) - (\Gamma_s+a)\lvert \tau_1-\tau_2\rvert}, 
\end{equation}
we conclude that each exponential gives rise to a large-$a$ pseudomode $d_{\beta-}$, labeled by the multi-index $\beta=(m,n,s)$,  and having frequency $\Omega_\beta-ia$, decay rate $\Gamma_\beta+a$, and characterized by the interaction Hamiltonian $H_{\text{int},\beta-} = \lambda_s^{\prime\prime*} S_m^\dagger d_{\alpha-}+\lambda_s^{\prime*} S_n^\dagger d_{\alpha-}^\dagger$. 

Ultimately, using the purification technique in Sec.~\ref{sec:PMadiabaticlimits}, we find the mapping 
\begin{equation}
[H_{\text{int},\alpha+},\boldsymbol{\cdot}] \to \lambda_s'd_{\alpha+}[S_n,\boldsymbol{\cdot}] + \lambda_s'' d_{\alpha+}^\dagger S_m[\boldsymbol{\cdot}],~~~[H_{\text{int},\beta-},\boldsymbol{\cdot}] \to \lambda_s^{\prime*}[S_n^\dagger,\boldsymbol{\cdot}]d_{\beta-}^\dagger - \lambda_s^{\prime\prime*}[\boldsymbol{\cdot}]S_m^\dagger d_{\beta-}, 
\end{equation}
yielding the same purified pseudomode interactions as those in Eq.~(12) in the main text. We have summarized these results in Table.~\ref{tab:sysdyn}. 

\subsection{Bath input-output}

Since the analysis of field superoperators largely overlaps with that in the previous subsection, we directly summarize the results for the simulation of a left field superoperator $\Phi_j^{l}(t_j)$ in Table.~\ref{tab:lfieldsuper}, and for the simulation of a right one $\Phi_j^r(t_j)$ in Table.~\ref{tab:rfieldsuper}.

\begin{table}[t]
\centering
\begin{tabular}{>{\centering}m{0.08\textwidth}>{\centering}m{0.295\textwidth}>{\centering}m{0.295\textwidth}>{\centering\arraybackslash}m{0.295\textwidth}}
\toprule
& \multicolumn{2}{|c|}{\textbf{Output}: $t_j=t\ge\tau$}  & \textbf{Input}: $t_j=0\le\tau$ \\
\midrule
\textbf{corr.} & $\langle \phi_j(t_j)X_n(\tau)\rangle_0\Theta(t_j-\tau)$  & $\langle X_n(\tau)\phi_j(t_j)\rangle_0\Theta(t_j-\tau)$ & $\langle X_n(\tau)\phi_j(t_j)\rangle_0\Theta(\tau-t_j)$  \\
\midrule
\textbf{decomp.} & $\sum_s w_s e^{-(i\Omega_s+\Gamma_s)(t_j-\tau)}$ & $\sum_s w_s e^{-(i\Omega_s-\Gamma_s)(\tau-t_j)}$ & $\sum_s w_s e^{-(i\Omega_s+\Gamma_s)(\tau-t_j)}$ \\
\midrule
\textbf{approx.} & $\sum_s w_s e^{-i(\Omega_s+ia)(t_j-\tau)-(\Gamma_s+a)\lvert t_j-\tau\rvert}$ & $\sum_s w_s e^{-i(\Omega_s-ia)(\tau-t_j)-(\Gamma_s+a)\lvert\tau-t_j\rvert}$ & $\sum_s w_s e^{-i(\Omega_s+ia)(\tau-t_j)-(\Gamma_s+a)\lvert \tau-t_j\rvert}$ \\
\midrule
\textbf{freq.} & $\Omega_s+ia$ & $\Omega_s-ia$ & $\Omega_s+ia$   \\
\midrule
\textbf{decay} & $\Gamma_s+a$ & $\Gamma_s+a$ & $\Gamma_s+a$ \\
\midrule
$(\mathbf{H_{\text{int}}},~\mathbf{\Phi}_\text{PM}^l)$ & $(\lambda_s' d_{\alpha+}^\dagger S_n,~\lambda_s'' d_{\alpha+})$, $\alpha=(j,n,s,+)$ & $(\lambda_s' d_{\beta-}S_n,~\lambda_s'' d_{\beta-}^\dagger)$, $\beta=(n,j,s,-)$ & $(\lambda_s' d_{{\alpha'}+}S_n,~\lambda_s'' d_{{\alpha'}+}^\dagger)$, $\alpha'=(n,j,s,+)$  \\
\midrule
\textbf{purif.} & $(\lambda_s'd_{\alpha+}^\dagger S_n[\boldsymbol{\cdot}],~\lambda_s''d_{\alpha+}^\dagger[\boldsymbol{\cdot}])$ & $(-\lambda_s'[\boldsymbol{\cdot}]S_nd_{\beta-},~\lambda_s''[\boldsymbol{\cdot}]d_{\beta-}^\dagger)$ & $(\lambda_sd_{\alpha'+}[S_n,\boldsymbol{\cdot}],~\lambda_sd_{\alpha'+}[\boldsymbol{\cdot}])$ \\
\bottomrule
\end{tabular}
\caption{This table contains the results for a left field superoperator $\Phi_j^{l}(t_j)$. The time arguments satisfy $0\le t_j,\tau \le t$. The last row corresponds to the purification of $[H_\text{int},\boldsymbol{\cdot}]$ and $\Phi_\text{PM}^l[\boldsymbol{\cdot}]$. In all rows, the constraint $w_s=\lambda_s'\lambda_s''$ should be imposed.  
}\label{tab:lfieldsuper}
\end{table}

\begin{table}[t]
\centering
\begin{tabular}{>{\centering}m{0.08\textwidth}>{\centering}m{0.295\textwidth}>{\centering}m{0.295\textwidth}>{\centering\arraybackslash}m{0.295\textwidth}}
\toprule
& \multicolumn{2}{|c|}{\textbf{Output}: $t_j=t\ge\tau$}  & \textbf{Input}: $t_j=0\le\tau$ \\
\midrule
\textbf{corr.} & $\langle \phi_j(t_j)X_n(\tau)\rangle_0\Theta(t_j-\tau)$  & $\langle X_n(\tau)\phi_j(t_j)\rangle_0\Theta(t_j-\tau)$ & $\langle \phi_j(t_j)X_n(\tau)\rangle_0\Theta(\tau-t_j)$  \\
\midrule
\textbf{decomp.} & $\sum_s w_s e^{-(i\Omega_s+\Gamma_s)(t_j-\tau)}$ & $\sum_s w_s e^{-(i\Omega_s-\Gamma_s)(\tau-t_j)}$ & $\sum_s w_s e^{-(i\Omega_s+\Gamma_s)(t_j-\tau)}$ \\
\midrule
\textbf{approx.} & $\sum_s w_s e^{-i(\Omega_s+ia)(t_j-\tau)-(\Gamma_s+a)\lvert t_j-\tau\rvert}$ & $\sum_s w_s e^{-i(\Omega_s-ia)(\tau-t_j)-(\Gamma_s+a)\lvert\tau-t_j\rvert}$ & $\sum_s w_s e^{-i(\Omega_s-ia)(t_j-\tau)-(\Gamma_s+a)\lvert t_j-\tau\rvert}$ \\
\midrule
\textbf{freq.} & $\Omega_s+ia$ & $\Omega_s-ia$ & $\Omega_s-ia$   \\
\midrule
\textbf{decay} & $\Gamma_s+a$ & $\Gamma_s+a$ & $\Gamma_s+a$ \\
\midrule
$(\mathbf{H_{\text{int}}},~\mathbf{\Phi}_\text{PM}^r)$ & $(\lambda_s' d_{\alpha+}^\dagger S_n,~\lambda_s'' d_{\alpha+})$, $\alpha=(j,n,s,+)$ & $(\lambda_s' d_{\beta-}S_n,~\lambda_s'' d_{\beta-}^\dagger)$, $\beta=(n,j,s,-)$ & $(\lambda_s' d_{{\beta'}-}^\dagger S_n,~\lambda_s'' d_{{\beta'}-})$, $\beta'=(j,n,s,-)$  \\
\midrule
\textbf{purif.} & $(\lambda_s'd_{\alpha+}^\dagger S_n[\boldsymbol{\cdot}],~\lambda_s''d_{\alpha+}^\dagger[\boldsymbol{\cdot}])$ & $(-\lambda_s'[\boldsymbol{\cdot}]S_nd_{\beta-},~\lambda_s''[\boldsymbol{\cdot}]d_{\beta-}^\dagger)$ & $(\lambda_sd_{\alpha'+}[S_n,\boldsymbol{\cdot}],~\lambda_sd_{\alpha'+}[\boldsymbol{\cdot}])$ \\
\bottomrule
\end{tabular}
\caption{This table contains the results for a right field superoperator $\Phi_j^{r}(t_j)$. The time arguments satisfy $0\le t_j,\tau \le t$. The last row corresponds to the purification of $[H_\text{int},\boldsymbol{\cdot}]$ and $[\boldsymbol{\cdot}]\Phi_\text{PM}^r$. In all rows, the constraint $w_s=\lambda_s'\lambda_s''$ should be imposed. 
}\label{tab:rfieldsuper}
\end{table}

\section{Relation to the Free-pole HEOM}\label{sec:relationtofpHEOM}

Here, we  show an explicit equivalence between Eq.~(9) in the main text and the free-pole HEOM reported in Ref.~[46], for the  evaluation of the reduced system dynamics for an open quantum system characterized by a single system coupling operator $S$, and a single bath coupling operator $X=\sum_k g_k(b_k+b_k^\dagger)$ in terms of the coupling constants $g_k$. To show this, we assume the decomposition $\langle X(\tau_2)X(\tau_1)\rangle_0 = \sum_sw_s\exp{[-(i\Omega_s+\Gamma_s)(\tau_2-\tau_1)]}$ for $\tau_2\ge\tau_1$, and write Eq.~(9) in the main text explicitly as
\begin{equation}\label{eq:fpheom_me}
\begin{array}{lll}
\displaystyle i\frac{d\rho}{dt} &\displaystyle = [H_\text{S},\rho] + \sum_s\left((\Omega_s-i\Gamma_s)d_{s+}^\dagger d_{s+}\rho + \sqrt{w_s} d_{s+}[S,\rho] + \sqrt{w_s}d_{s+}^\dagger S\rho \right) \\
&\displaystyle~~~ - \sum_s\left( (\Omega_s+i\Gamma_s)\rho d_{s-}^\dagger d_{s-} - \sqrt{w_s^*}[S,\rho]d_{s-}^\dagger + \sqrt{w_s^*}\rho Sd_{s-} \right). 
\end{array}
\end{equation}

Now, the auxiliary density matrices (ADOs) in the HEOM can be identified as the matrices $\rho_{\mathbf{m},\mathbf{n}} = \langle\mathbf{m}|\rho|\mathbf{n}\rangle$, where $|\mathbf{m}\rangle$, $|\mathbf{n}\rangle$ denote the Fock states of the purified pseudomodes, with the occupations $m_s,~n_s$ of the modes $d_{s\pm}$ given by the sequences $\mathbf{m}=(m_1,m_2,\cdots)$ and $\mathbf{n}=(n_1,n_2,\cdots)$. Note that, these ADOs have the same dimension as the reduced density matrix $\rho_\text{S}$. From Eq.~(\ref{eq:fpheom_me}), the ADOs $\rho_{\mathbf{m},\mathbf{n}}$ obey the equations of motion 
\begin{equation}\label{eq:fpheom}
\begin{array}{lll}
\displaystyle i\frac{d\rho_{\mathbf{m},\mathbf{n}}}{dt} &\displaystyle = [H_\text{S},\rho_{\mathbf{m},\mathbf{n}}] + \sum_s\left((\Omega_s-i\Gamma_s)m_s d_{s+}\rho_{\mathbf{m},\mathbf{n}} + \sqrt{(m_s+1)w_s}[S,\rho_{\mathbf{m}_s^+,\mathbf{n}}] + \sqrt{m_sw_s} S\rho_{\mathbf{m}_s^-,\mathbf{n}} \right) \\
&\displaystyle~~~ - \sum_s\left( (\Omega_s+i\Gamma_s)n_s\rho_{\mathbf{m},\mathbf{n}}  - \sqrt{(n_s+1)w_s^*}[S,\rho_{\mathbf{m},\mathbf{n}_s^+}] + \sqrt{n_sw_s^*}\rho_{\mathbf{m},\mathbf{n}_s^-}S \right), 
\end{array}
\end{equation}
with $\mathbf{m}_s^\pm = (m_1,\cdots, m_s\pm1,\cdots)$ and $\mathbf{n}_s^\pm = (n_1,\cdots, n_s\pm1,\cdots)$. The expression in Eq.~(\ref{eq:fpheom}) is exactly the same as the free-pole HEOM established in Ref.~[46], cf., Eq.~(5) in the supplemental material of Ref.~[46].

\section{The Coupled-cavity Waveguide Model}\label{sec:coupledcavityQED}

For completeness, in this section, we summarize the description of the coupled-cavity waveguide model considered in the main text, which has been  experimentally realized in Ref.~[91] and theoretically studied in Ref.~[90]. The model consists of two optical cavities, labeled as $C_{n}$ with $n=1,2$, side coupled to a linear waveguide. Each cavity further couples to a two-level emitter $E_n$ made of quantum dot (QD), whose excitonic energy can be controlled by tuning the temperature of the phonon environment experienced by the QD~[91]. The cavity and waveguide free Hamiltonian are $H_\text{c}=\sum_{n=1,2}\omega_{c,n}c_n^\dagger c_n$ and $H_\text{wg} = \int dk \omega_k b_k^\dagger b_k$ respectively, where $k$ denotes the quantum number of waveguide eigenmodes, e.g., the wavenumber of plane waves in a one-dimensional waveguide, and $\omega_{c,n}$ denotes resonance frequencies of the two cavities. The dispersion of the waveguide is $\omega_k = v_g\lvert k\rvert$ with $v_g$ the group velocity of light. The bare Hamiltonian of the two emitters is $H_\text{e} = \sum_{n=1,2}\omega_{e,n}\sigma_n^+\sigma_n^-$ with $\omega_{e,n}$ the bare emitter excitation energies. The cavities interact with the waveguide via $H_\text{c,wg}=\sum_{n}\int dk g_n\sqrt{\lvert k\rvert}(e^{-ikx_n}c_nb_k^\dagger + \text{H.c.})$, which we assume is weak such that the rotating wave approximation (RWA) is valid. Note that, since the two cavities couple at distinct spatial locations $x_{1,2}$ of the waveguide, their coupling amplitudes carry spatial information in terms of the phases $e^{-ikx_n}$. The whole system is thus described by the Hamiltonian
\begin{equation}
H = H_\text{e} + H_\text{c} + H_\text{ec} + H_\text{wg} + H_\text{c,wg}, 
\end{equation}
where the emitter-cavity interaction can be the Jannes-Cummings (JC) interaction $H_\text{ec} = \sum_n (\sigma_n^+X_n+\text{H.c.})$ with $X_n=V_nc_n^\dagger$, or the Rabi interaction $H_\text{ec} = \sum_n \sigma_n^xX_n$ with $X_n=V_nc_n^\dagger+\text{H.c.}$. Here without loss of generality we always assume real-valued $V_n$. 

In our approach, we take an open quantum system point of view, and treat the two cavities and the waveguide together as a Gaussian environment of the emitters, i.e., $H_\text{B} = H_\text{c}+H_\text{wg}+H_\text{c,wg}$. In turn, this means that the emitter dynamics is determined by four free bath correlations $\langle c_n(t)c_m^\dagger(0)\rangle_0 = \text{Tr}[e^{iH_\text{B}t}c_n e^{-iH_\text{B}t}c_m^\dagger\rho_0]$ for both Jaynes-Cummings and Rabi interactions. In line with Ref.~[91], we use the global vacuum state $\rho_0 = |0,0,\mathbf{0}\rangle\langle0,0,\mathbf{0}|$ of the cavities and waveguide as the bath reference state (upon which defining more general initial conditions), where $|n_1,n_2,\mathbf{m}\rangle$ denote the Fock state with $n_1,~n_2,~\mathbf{m}=(m_{k_1},m_{k_2},\cdots)$ representing occupation numbers of the two cavities and waveguide modes, respectively. Note that, the invariance of the chosen bath initial state under free bath dynamics, i.e., $[\rho_0,H_\text{B}]=0$, implies the time translational invariance of the correlation functions, i.e., $\langle c_n(t_2)c_m(t_1)\rangle_0 = \langle c_n(t_2-t_1)c_m(0)\rangle_0$. 

Below we present detailed calculations of $\langle c_n(t)c_m^\dagger(0)\rangle_0$. 

\subsection{Bath Free Correlations}

As a result of the excitation number conservation of the free bath Hamiltonian $H_\text{B}$, $\langle c_n(t)c_m^\dagger(0)\rangle_0$ can be shown to be equivalent to the amplitudes of the  wavefunction in the single-excitation subspace. For example, for $\langle c_1(t)c_1^\dagger(0)\rangle_0$, we have $\langle c_1(t)c_1^\dagger(0)\rangle_0 = \langle 0,0,\mathbf{0}|e^{iH_\text{B}t}c_1e^{-iH_\text{B}t}c_1^\dagger|0,0,\mathbf{0}\rangle = \langle 1,0,\mathbf{0}|\psi(t)\rangle$ where $|\psi(t)\rangle = e^{-iH_\text{B}t}|0,0,\mathbf{0}\rangle$ is the wavefunction in the single-excitation subspace for the initial state $|1,0,\mathbf{0}\rangle$. Now we solve the Schr\"odinger equation in the single-excitation subspace to find the solution of $|\psi(t)\rangle$. 

We first expand $|\psi(t)\rangle$ in the Fock basis 
\begin{equation}
|\psi(t)\rangle = \epsilon_1|1,0,\mathbf{0}\rangle + \epsilon_2|0,1,\mathbf{0}\rangle + \int dk \beta_k|0,0,1_k\rangle, 
\end{equation}
where $\epsilon_1,~\epsilon_2,~\beta_k$ denote the complex amplitudes. Substituting this back into the Schr\"odinger equation $id|\psi(t)\rangle/dt = H_\text{B}|\psi(t)\rangle$, we obtain a set of coupled first-order linear differential equations
\begin{equation}\label{eq:couplede}
\begin{array}{lll}
\displaystyle i\frac{d\epsilon_1(t)}{dt} &\displaystyle = \omega_{c,1}\epsilon_1(t) + g_1\int dk \sqrt{\lvert k\rvert} e^{ikx_1}\beta_k, \\
\displaystyle i\frac{d\epsilon_2(t)}{dt} &\displaystyle = \omega_{c,2}\epsilon_2(t) + g_2\int dk \sqrt{\lvert k\rvert} e^{ikx_2}\beta_k, \\
\displaystyle i\frac{d\beta_k(t)}{dt} &\displaystyle = \omega_k\beta_k(t) + g_1\sqrt{\lvert k\rvert} e^{-ikx_1}\epsilon_1 + g_2\sqrt{\lvert k\rvert} e^{-ikx_2}\epsilon_2. 
\end{array}
\end{equation}
The formal solution of the last equation is
\begin{equation}
\beta_k(t) = e^{-i\omega_k t}\beta_k(0) - i\int_0^td\tau\Big[g_1\sqrt{\lvert k\rvert}e^{-ikx_1}e^{i\omega_k(\tau-t)}\epsilon_1(\tau) + g_2\sqrt{\lvert k\rvert}e^{-ikx_2}e^{i\omega_k(\tau-t)}\epsilon_2(\tau) \Big]. 
\end{equation}
Substituting this solution into the first line of Eq.~(\ref{eq:couplede}), we obtain 
\begin{equation}
\begin{array}{lll}
\displaystyle i\frac{d\epsilon_1(t)}{dt} &\displaystyle = \omega_{c,1} \epsilon_1(t)  + g_1\int dk\sqrt{\lvert k\rvert} e^{ikx_1}e^{-i\omega_kt}\beta_k(0) \\
&\displaystyle~~~ - i\int dk \lvert k\rvert e^{ikx_1} \int_0^td\tau\Big[g_1^2e^{-ikx_1}e^{i\omega_k(\tau-t)}\epsilon_1(\tau) + g_1g_2e^{-ikx_2}e^{i\omega_k(\tau-t)}\epsilon_2(\tau) \Big] \\
&\displaystyle = \omega_{c,1} \epsilon_1(t)  +  g_1\int dk\sqrt{\lvert k\rvert} e^{ikx_1}e^{-i\omega_kt}\beta_k(0) \\
&\displaystyle~~~ - i\frac{1}{v^2}\int_0^td\tau\int_0^\infty d\omega_k\omega_k \Big[2g_1^2e^{i\omega_k(\tau-t)}\epsilon_1(\tau) + 2g_1g_2\cos\left[ \frac{\omega_k(x_1-x_2)}{v}\right]e^{i\omega_k(\tau-t)}\epsilon_2(\tau) \Big]. 
\end{array}
\end{equation}

In typical optical experiments, the cavity frequencies $\omega_{c,n}$ are much larger than other frequency scales, so the cavity amplitudes $\epsilon_n(t)$ contain a fast oscillating phase $e^{-i\omega_{c,n}t}$. In order to incorporate this feature, we introduce a reference frequency $\omega_0$ which we assume is close to bare cavity and atomic frequencies, meaning that all relevant physical processes occur around $\omega_0$. We move to the rotating frame oscillating with frequency $\omega_0$ by defining $\epsilon_n(t) = e^{-i\omega_0t}\epsilon_n'(t)$, such that
\begin{equation}
\begin{array}{lll}
\displaystyle i\frac{d\epsilon_1'(t)}{dt} &\displaystyle = \delta\omega_{c,1}\epsilon_1'(t) + g_1\int dk\sqrt{\lvert k\rvert} e^{ikx_1}e^{-i(\omega_k-\omega_0)t}\beta_k(0) \\
&\displaystyle~~~ - i\frac{1}{v^2}\int_0^td\tau\int_0^\infty d\omega_k\omega_k \Big[2g_1^2e^{i(\omega_k-\omega_0)(\tau-t)}\epsilon_1'(\tau) + 2g_1g_2\cos \left[\frac{\omega_k(x_1-x_2)}{v}\right]e^{i(\omega_k-\omega_0)(\tau-t)}\epsilon_2'(\tau) \Big], 
\end{array}
\end{equation}
where $\delta\omega_{c,1}=\omega_{c,1}-\omega_0$ is the cavity detuning. In line with the Weisskopf-Wigner theory, since all physical processes occur around a narrow frequency window around $\omega_0$, we can simply replace the linear spectral density of the waveguide by the reference frequency, i.e., $\omega_k\to\omega_0$. This allows to perform a change of variable $\omega_k\to\omega_k-\omega_0$ in the above integral and accordingly extend the lower bound of the integral to $-\infty$, which leads to
\begin{equation}
\begin{array}{lll}
\displaystyle i\frac{d\epsilon_1'(t)}{dt} &\displaystyle = \delta\omega_{c,1}\epsilon_1'(t) +  g_1\int dk\sqrt{\lvert k\rvert} e^{ikx_1}e^{-i(\omega_k-\omega_0)t}\beta_k(0) \\
&\displaystyle~~~ - i\frac{2\pi\omega_0}{v^2}\int_0^t d\tau\Bigg[2g_1^2\delta(\tau-t)\epsilon_1'(\tau) + g_1g_2e^{-i\omega_0d/v}\delta\left(\frac{x_1-x_2}{v}+\tau-t\right)\epsilon_2'(\tau) \\
&\displaystyle~~~ + g_1g_2e^{i\omega_0d/v}\delta\left(-\frac{x_1-x_2}{v}+\tau-t\right)\epsilon_2'(\tau) \Bigg] \\ 
&\displaystyle = \delta\omega_{c,1}\epsilon_1'(t) + g_1\int dk\sqrt{\lvert k\rvert} e^{ikx_1}e^{-i(\omega_k-\omega_0)t}\beta_k(0)  - i \frac{\kappa_1}{2}\epsilon_1'(t) - i \frac{\sqrt{\kappa_1\kappa_2}}{2}e^{i\omega_0t_d} \epsilon_2'(t-t_d), 
\end{array}
\end{equation}
where we have defined the cavity dissipation rate $\kappa_{1(2)} = 4\pi g_{1(2)}^2\omega_0/v^2$, the distance between the two cavities $d=x_2-x_1\ge0$ and the traveling time $t_d=d/v$. Similarly, for $\epsilon_2$ we obtain
\begin{equation}
i\frac{d\epsilon_2'(t)}{dt} = \delta\omega_{c,2} \epsilon_2'(t)  +  g_2\int dk\sqrt{\lvert k\rvert} e^{ikx_2}e^{-i(\omega_k-\omega_0)t}\beta_k(0)  - i \frac{\kappa_2}{2}\epsilon_2'(t) - i \frac{\sqrt{\kappa_1\kappa_2}}{2} e^{i\omega_0t_d}\epsilon_1'(t-t_d). 
\end{equation}

We recall that at $t=0$, $\beta_k=0$, which allows to further simplify these equations as
\begin{equation}
\frac{d\epsilon_1'(t)}{dt} = -\left(i\delta\omega_{c,1}+ \frac{\kappa_1}{2}\right)\epsilon_1'(t) - \frac{\sqrt{\kappa_1\kappa_2}}{2}e^{i\omega_0t_d}\epsilon_{2}'(t-t_d),~\frac{d\epsilon_2(t)}{dt} = -\left(i\omega_{c,2}+ \frac{\kappa_2}{2}\right)\epsilon_2(t) - \frac{\kappa_{21}}{2} e^{i\omega_0t_d}\epsilon_{1}(t-t_d). 
\end{equation}
Now we employ the Laplace transform $\epsilon_n'[s]=\int_0^\infty dt \epsilon_n'(t)e^{-st}$ to obtain
\begin{equation}
\begin{array}{lll}
s\epsilon_1'[s]-\epsilon_1'(0) &\displaystyle = -\left(i\delta\omega_{c,1}+ \frac{\kappa_1}{2}\right)\epsilon_1'[s] - \frac{\sqrt{\kappa_1\kappa_2}}{2}e^{i\omega_0t_d} e^{-st_d}\epsilon_2'[s], \\
s\epsilon_2'[s]-\epsilon_2'(0) &\displaystyle = -\left(i\delta\omega_{c,2}+ \frac{\kappa_2}{2}\right)\epsilon_2'[s] - \frac{\sqrt{\kappa_1\kappa_2}}{2}e^{i\omega_0t_d} e^{-st_d}\epsilon_1'[s],
\end{array}
\end{equation}
which can be solved to yield the solutions
\begin{equation}\label{eq:solslaplace}
 \epsilon_1'[s]  = \frac{\epsilon_1'(0)-\frac{\sqrt{\kappa_1\kappa_2}}{2}e^{-st_d}e^{i\omega_0t_d}\frac{1}{s+(i\delta\omega_{c,2}+{\kappa_2}/{2})}\epsilon_2'(0)}{s+(i\delta\omega_{c,1}+{\kappa_1}/{2})-\frac{\kappa_1\kappa_2}{4}e^{2i\omega_0t_d}\frac{e^{-2st_d}}{s+(i\delta\omega_{c,2}+{\kappa_2}/{2})}}, 
 \epsilon_2'[s]  = \frac{\epsilon_2'(0)-\frac{\sqrt{\kappa_1\kappa_2}}{2}e^{-st_d}e^{i\omega_0t_d}\frac{1}{s+(i\delta\omega_{c,1}+{\kappa_1}/{2})}\epsilon_1'(0)}{s+(i\delta\omega_{c,2}+{\kappa_2}/{2})-\frac{\kappa_1\kappa_2}{4}e^{2i\omega_0t_d}\frac{e^{-2st_d}}{s+(i\delta\omega_{c,1}+{\kappa_1}/{2})}}, 
\end{equation}

Below we consider two different regimes separately. 

\subsubsection{Case I: Negligible Delay Time}

First we consider the case of negligible delay time, i.e., $\kappa_{1(2)}t_d\ll1$ but the phase $\omega_0t_d$ is not negligible. In other words, this means that the wavelength $\lambda = 2\pi v_g/\omega_0$ is comparable to distance $d$ between the emitters, i.e., $\lambda\sim d$. Note that, most state-of-the-art experiments are in this regime. Since Eq.~(\ref{eq:solslaplace}) describes the slow dynamics, we can safely approximate $\exp{[-2st_d]}\approx 1$. Then the poles of $\epsilon_n'[s]$ are determined by the quadratic algebraic equation 
\begin{equation}
\Big[s+\left(i\delta\omega_{c,1}+\frac{\kappa_1}{2}\right)\Big]\Big[s+\left(i\delta\omega_{c,2}+\frac{\kappa_2}{2}\right)\Big]-\frac{\kappa_1\kappa_2}{4}e^{2i\omega_0t_d} = 0, 
\end{equation}
which yields the analytical solutions
\begin{equation}
s_{1,2} = -\frac{i(\delta\omega_{c,1}+\delta\omega_{c,2})+\kappa_1/2+\kappa_2/2}{2} \pm \frac12\sqrt{\Big(i\delta\omega_{c,1}-i\delta\omega_{c,2}+\frac{\kappa_1}{2}-\frac{\kappa_2}{2}\Big)^2+\kappa_1\kappa_2e^{2i\omega_0t_d}}. 
\end{equation}
In order to calculate $\langle c_1(t)c_1^\dagger(0)\rangle_0$ and $\langle c_2(t)c_1^\dagger(0)\rangle_0$, we use the initial conditions $\epsilon_1'(0)=1,~\epsilon_2'(0)=0$ and do the inverse Laplace transform to obtain
\begin{equation}\label{eq:corr11_smalldelay}
\renewcommand{\arraystretch}{3}{
\begin{array}{lll}
\langle c_1(t)c_1^\dagger(0)\rangle_0 &\displaystyle = e^{-i\omega_0t}L^{-1}(\epsilon_1'[s]) = e^{-i\omega_0t}L^{-1}\left(\frac{s+i\delta\omega_{c,2}+\kappa_2/2}{(s-s_1)(s-s_2)}\right) = w_{11,1}e^{-i\omega_0t+s_1t}+w_{11,2}e^{-i\omega_0t+s_2t},  \\
\langle c_2(t)c_1^\dagger(0)\rangle_0 &\displaystyle= e^{-i\omega_0t}L^{-1}(\epsilon_2'[s]) = e^{-i\omega_0t}L^{-1}\left(\frac{-\frac{\sqrt{\kappa_1\kappa_2}}{2}e^{i\omega_0t_d}}{(s-s_1)(s-s_2)}\right) = w_{21,1}e^{-i\omega_0t+s_1t}+w_{21,2}e^{-i\omega_0t+s_2t},  
\end{array}}
\end{equation}
for $t\ge0$, with the coefficients defined as
\begin{equation}
w_{11,1} = \frac{s_1+i\delta\omega_{c,2}+\kappa_2/2}{s_1-s_2},~w_{11,2}=\frac{s_2+i\delta\omega_{c,2}+\kappa_2/2}{s_2-s_1},~w_{21,1} = \frac{-\frac{\sqrt{\kappa_1\kappa_2}}{2}e^{i\omega_0t_d}}{s_1-s_2},~w_{21,2}=\frac{-\frac{\sqrt{\kappa_1\kappa_2}}{2}e^{i\omega_0t_d}}{s_2-s_1}. 
\end{equation}

Similarly, for $\langle c_2(t)c_2^\dagger(0)\rangle_0$ and $\langle c_1(t)c_2^\dagger(0)\rangle_0$, we use the initial condition $\epsilon_1'(0)=0,~\epsilon_2'(0)=1$, and obtain
\begin{equation}\label{eq:corr22_smalldelay}
\renewcommand{\arraystretch}{3}{
\begin{array}{lll}
 \langle c_2(t)c_2^\dagger(0)\rangle_0 &\displaystyle = e^{-i\omega_0t}L^{-1}(\epsilon_2'[s]) = e^{-i\omega_0t}L^{-1}\left(\frac{s+i\delta\omega_{c,1}+\kappa_1/2}{(s-s_1)(s-s_2)}\right) = w_{22,1}e^{-i\omega_0t+s_1t}+w_{22,2}e^{-i\omega_0t+s_2t},  \\
 \langle c_1(t)c_2^\dagger(0)\rangle_0 &\displaystyle = e^{-i\omega_0t}L^{-1}(\epsilon_1'[s]) = e^{-i\omega_0t}L^{-1}\left(\frac{-\frac{\sqrt{\kappa_1\kappa_2}}{2}e^{i\omega_0t_d}}{(s-s_1)(s-s_2)}\right) = w_{12,1}e^{-i\omega_0t+s_1t}+w_{12,2}e^{-i\omega_0t+s_2t},   
\end{array}}
\end{equation}
for $t\ge0$, with the coefficients
\begin{equation}
w_{22,1} = \frac{s_1+i\delta\omega_{c,1}+\kappa_1/2}{s_1-s_2},~w_{22,2}=\frac{s_2+i\delta\omega_{c,1}+\kappa_1/2}{s_2-s_1},~w_{12,1} = \frac{-\frac{\sqrt{\kappa_1\kappa_2}}{2}e^{i\omega_0t_d}}{s_1-s_2},~w_{12,2}=\frac{-\frac{\sqrt{\kappa_1\kappa_2}}{2}e^{i\omega_0t_d}}{s_2-s_1}. 
\end{equation}

Lastly, for $t<0$, the correlations can be directly obtained from the time-reversal property $\langle c_n(t)c_m^\dagger(0)\rangle_0 = \langle c_m(-t)c_n^\dagger(0)\rangle_0^*$. 

We note that in this case, the emitter-cavity dynamics is described by a Lindblad master equation~[91]
\begin{equation}
\begin{array}{lll}
\displaystyle \frac{d\rho}{dt} &\displaystyle = -i[H_\text{e}'+H_\text{c}'+H_\text{ec},\rho] + \sum_{n=1,2}\frac{\kappa_n}{2}(2c_n\rho c_n^\dagger - c_n^\dagger c_n\rho - \rho c_n^\dagger c_n) \\
&\displaystyle~~~  + \frac{\sqrt{\kappa_1\kappa_2}}{2}\sum_{n\neq m}\Big[e^{-i\omega_0t_d}\big(c_m\rho c_n^\dagger - \rho c_n^\dagger c_m\big) + e^{i\omega_0t_d}\big(c_n\rho c_m^\dagger - c_m^\dagger c_n\rho \big)\Big],  
\end{array}
\end{equation}
where $H_\text{e}',~H_\text{c}'$ are the detuned emitter and cavity Hamiltonian with respect to the reference frequency $\omega_0$. For more details we refer to Ref.~[91] and references therein.

\subsubsection{Case II: Finite Delay Time}

\begin{figure}
\includegraphics[clip,width=8.5cm]{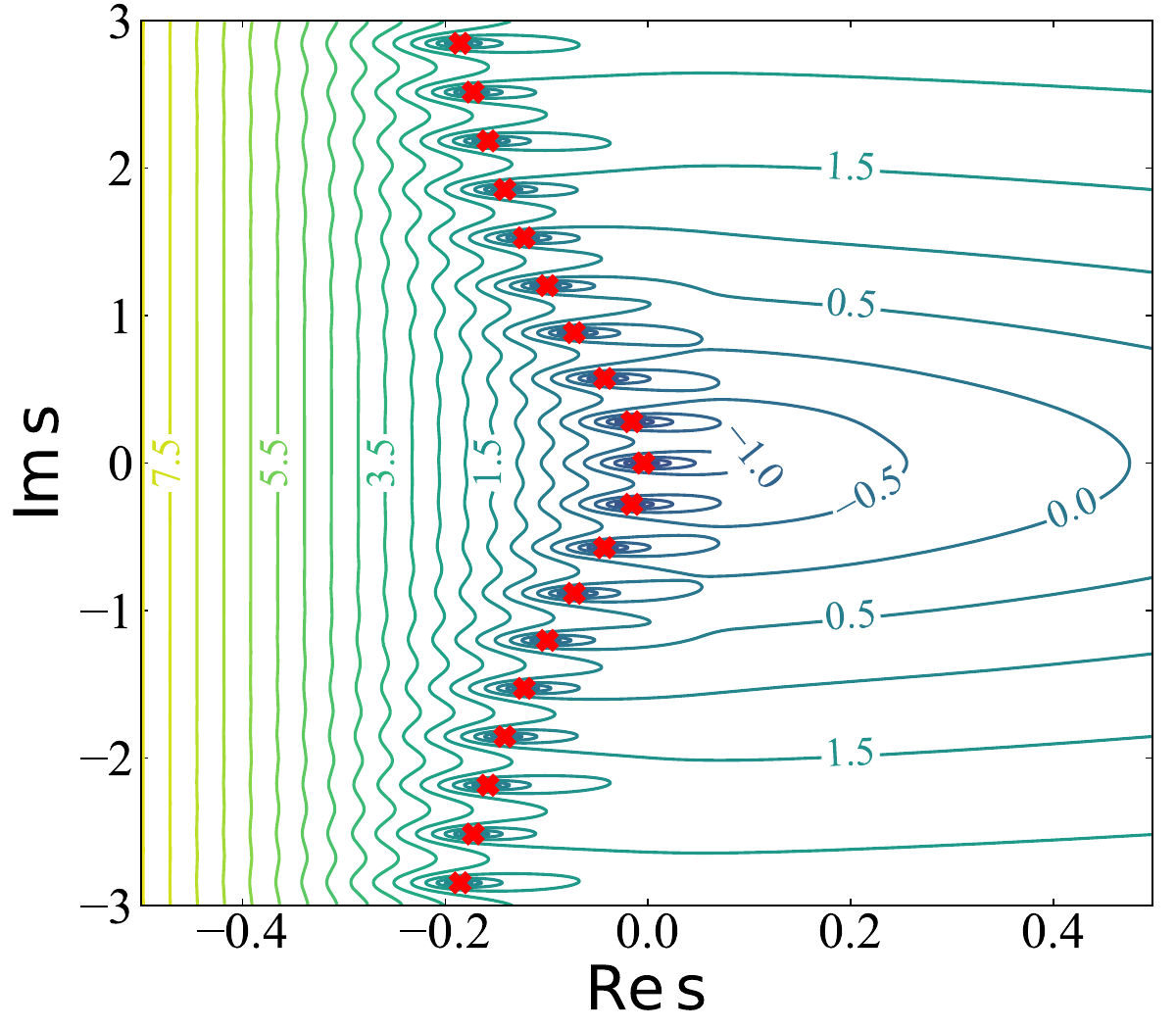}%
\caption{Contour plot of the function $\log\lvert F(s)\rvert$ in the complex $s$ plane. Red crosses mark the positions of the roots of $F(s)$. The parameters used here are the same as those in Fig.~3(b) and (c) in the main text for $x_d=1500\lambda_0$. 
}\label{fig:polesContour}
\end{figure}

\begin{figure}
\includegraphics[clip,width=16.5cm]{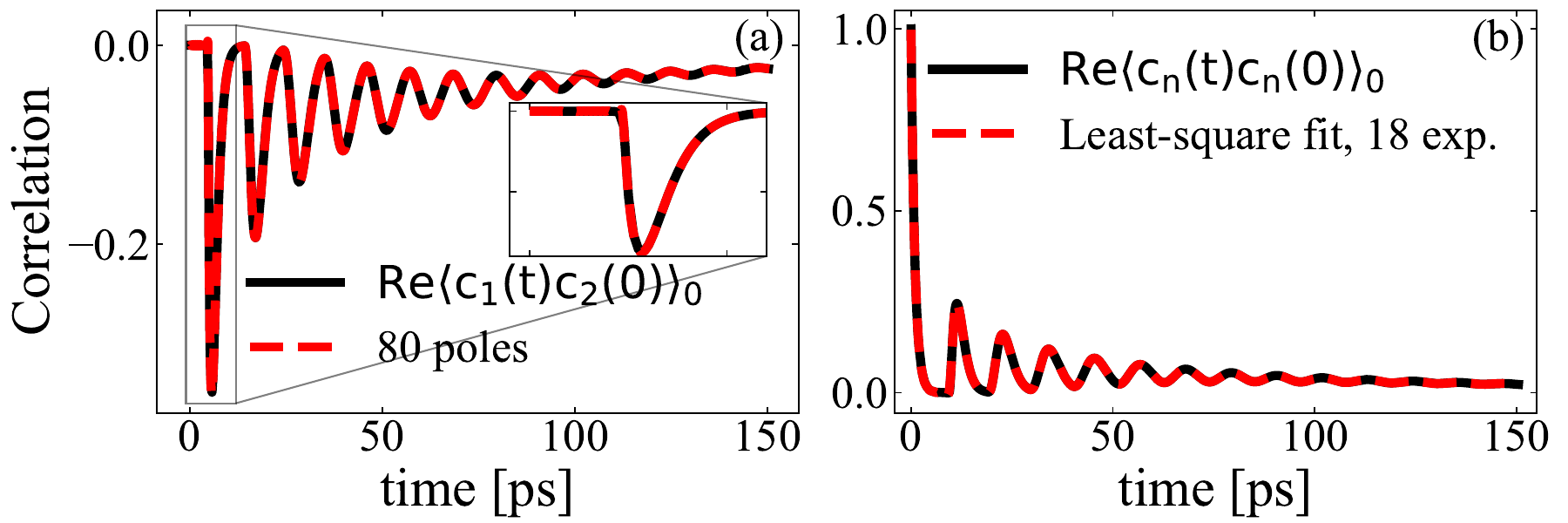}%
\caption{Bath correlations (black solid) and their exponential decomposition (red dashed) obtained (a) by using $80$ roots of $F(s)$ for $\langle c_1(t)c_2^\dagger(0)\rangle_0$ and, (b) by performing a least-square fit with $18$ exponentials for $\langle c_n(t)c_n^\dagger(0)\rangle_0$. The inset in (a) highlights the vanishing of $\langle c_1(t)c_2^\dagger(0)\rangle_0$ before $t\approx\SI{7}{\pico\second}$, which is well captured by the reconstructed correlation (red dashed) with $80$ poles. All correlations are real-valued, so only their real part is plotted. Parameters here are the same as those in Fig.~\ref{fig:polesContour}. 
}\label{fig:BathCorrLargeDelay}
\end{figure}

When the delay time is not negligible, i.e., $\kappa_{1(2)}t_d\sim1$, the poles of $\epsilon_n'[s]$ are determined by the transcendental equation, 
\begin{equation}\label{eq:Feq}
F(s) \equiv \Big[s+\left(i\delta\omega_{c,1}+\frac{\kappa_1}{2}\right)\Big]\Big[s+\left(i\delta\omega_{c,2}+\frac{\kappa_2}{2}\right)\Big]-\frac{\kappa_1\kappa_2}{4}e^{2i\omega_0t_d} e^{-2st_d} = 0, 
\end{equation}
which has no analytical solutions. However, its numerical solutions can readily be found using standard routines to search for the roots of $\lvert F(s)\rvert$. In Fig.~\ref{fig:polesContour}, we plot the function $\log\lvert F(s)\rvert$ using the parameters in Fig.~3(b) and (c) in the main text for $x_d=1500\lambda_0$, and mark all roots (red crosses) in the range $-0.45\le\text{Re}s\le0.45, -3\le\text{Im}s\le3$. For large $t_d$, the poles gradually form a quasi-continuum, which is necessary to correctly reproduce the vanishing of the cross-correlations $\langle c_1(t)c_2^\dagger(0)\rangle_0$ or $\langle c_2(t)c_1^\dagger(0)\rangle_0$ before $t\approx\SI{7}{\pico\second}$, as illustrated in Fig.~\ref{fig:BathCorrLargeDelay}(a). Note that for the two less ``singular" self-correlations $\langle c_n(t)c_n^\dagger(0)\rangle_0$, we  have used least-square fitting to generate a more efficient exponential decomposition, as shown in Fig.~\ref{fig:BathCorrLargeDelay}(b).

\subsection{Model Parameters and Number of Purified Pseudomodes Used in Numerical Simulations}

In Table.~\ref{tab:params}, we summarize the model parameters used in our numerics to produce Fig.~3 in the main text. The chosen parameters are consistent with the experiment conducted in Ref.~[91].

\begin{table}[t]
\centering
\begin{tabular}{>{\centering}m{0.1\textwidth}>{\centering}m{0.18\textwidth}>{\centering}m{0.18\textwidth}>{\centering}m{0.08\textwidth}>{\centering}m{0.08\textwidth}>{\centering}m{0.08\textwidth}>{\centering\arraybackslash}m{0.13\textwidth}}
\toprule
 & $(\omega_{c,1},\omega_{c,2})$  & $(\omega_{e,1},\omega_{e,2})$ & $Q_n$ & $\kappa_n$ & $\kappa_{i,n}$ & $(V_1,V_2)$ \\
\toprule
\textbf{Fig.3(a-c)} & $(\SI{945.0}{\nano\metre},\SI{945.0}{\nano\metre})$ & $(\SI{945.0}{\nano\metre},\SI{945.0}{\nano\metre})$ & $10^3$ & $\omega_{c,n}/Q_n$ & $\kappa_{n}/20$ & $(\kappa_1/10,\kappa_2/5)$   \\
\midrule
\textbf{Fig.3(d)} & $(\SI{945.0}{\nano\metre},\SI{945.76}{\nano\metre})$ & $(\SI{944.62}{\nano\metre},\SI{945.38}{\nano\metre})$ & $10^3$ & $\omega_{c,n}/Q_n$ & $\kappa_{n}/20$ & $(\kappa_1/10,\kappa_2/5)$   \\
\bottomrule
\end{tabular}
\caption{This table contains model parameters used in the numerical simulations.
}\label{tab:params}
\end{table}

We also summarize the number of modes needed to obtain the converged results shown in Fig.~3 in the main text. Specifically, for $x_d=1500\lambda_0$ [Fig.~3(b) and (c)], $396$ purified pseudomodes are used to simulate the reduced system dynamics, among which each postive- or negative-time contribution of the self-correlations $\langle c_n(t)c_n^\dagger(0)\rangle_0$ costs $18$ modes while that of the cross-correlations ($\langle c_1(t)c_2^\dagger(0)\rangle_0$ or $\langle c_2(t)c_1^\dagger(0)\rangle_0$) costs $80$ modes. No extra modes are required to compute cavity occupations since they can be obtained from the modes that model the cross-correlations. For $x_d=4\lambda_0$ [Fig.~3(b), (d) and (e)], $16$ modes are used for the reduced system dynamics with $2$ for each positive- or negative-time contribution. In addition, in Fig.~3(d) $8$ extra modes are required to model the bath non-Gaussian initial state $\rho_\text{B}=c_1^\dagger|0\rangle\langle0|c_1$.